\documentclass[final,12p,times,twocolumn]{elsarticle}
\usepackage{lineno,hyperref}
\usepackage{graphicx}
\usepackage[linesnumbered,ruled,lined]{algorithm2e}
\usepackage{float}
\usepackage{tikz}
\usepackage{epstopdf}
\usepackage{amssymb}
\usepackage{amsmath}
\usepackage{listings}
\usepackage{tikz}
\usepackage{subcaption}
\usepackage{rotating}
\journal{Comput. Phys. Commun }
\newcommand{\tr}{\mathop{\mathrm{tr}}}

\newcommand{\br}{\mathbf{r}}
\makeatletter
\@addtoreset{table}{section}
\makeatother

\makeatletter
\@addtoreset{figure}{section}
\makeatletter

\begin{document}
%
%
\begin{frontmatter}
		\title{SPHinXsys: an open-source multi-physics and multi-resolution library based on smoothed particle  hydrodynamics}
		\author[myfirstaddress]{Chi Zhang }
		\ead{c.zhang@tum.de}
		\author[myfirstaddress]{Massoud Rezavand}
		\ead{massoud.rezavand@tum.de}
		\author[myfirstaddress]{Yujie Zhu}
		\ead{yujie.zhu@tum.de}
		\author[mysecondaryaddress]{Yongchuan Yu}
		\ead{yongchuan.yu@tum.de}
		\author[myfirstaddress]{Dong Wu}
		\ead{dong.wu@tum.de}
		\author[myfirstaddress]{Wenbin Zhang}
		\ead{wenbin.zhang@tum.de}
		\author[myfirstaddress]{Jianhang Wang}
		\ead{jianhang.wang@tum.de}
		\author[myfirstaddress]{Xiangyu Hu \corref{mycorrespondingauthor}}
		\ead{xiangyu.hu@tum.de}
		\address[myfirstaddress]{Department of Mechanical Engineering, 
				Technical University of Munich, 85748 Garching, Germany}
		\address[mysecondaryaddress]{Department of Aerospace and Geodesy, 
			Technical University of Munich, 82024 Taufkirchen, Germany}
		\cortext[mycorrespondingauthor]{Corresponding author.}
\begin{abstract}
In this paper, 
we present an open-source multi-resolution and multi-physics library: 
SPHinXsys (pronunciation: s'finksis) which is an acronym for 
\underline{S}moothed \underline{P}article \underline{H}ydrodynamics (SPH)
for \underline{in}dustrial comple\underline{X} \underline{sys}tems.
As an open-source library, 
SPHinXsys is developed and released under the terms of Apache License (2.0). 
Along with the source code, 
a complete documentation is also distributed to make the compilation and execution easy. 
SPHinXsys aims at modeling coupled multi-physics industrial dynamic systems including fluids, 
solids, multi-body dynamics and beyond, in a multi-resolution unified SPH framework.
As an SPH solver, SPHinXsys has many advantages namely,
(1) the generic design provides a C++ API  showing a very good flexibility when building domain-specific applications, 
(2) numerous industrial or scientific applications can be coupled within the same framework and 
(3) with the open-source philosophy, the community of users can collaborate and improve the library.
SPHinXsys presently (v0.2.0) includes validations and applications in the fields of fluid dynamics, 
solid dynamics, 
thermal and mass diffusion, reaction-diffusion, electromechanics and 
fluid-structure interactions (FSI).
\end{abstract}
\end{frontmatter}
%
%
{\bf PROGRAM SUMMARY}

\begin{small}
	\noindent
	{\em Program Title:}  SPHinXsys                                    \\
	{\em Current Version:}  v0.2.0                                   \\
	{\em CPC Library link to program files:} (to be added by Technical Editor) \\
	{\em Repository link:}  https://github.com/Xiangyu-Hu/SPHinXsys \\
	{\em Code Ocean capsule:} https://doi.org/10.24433/CO.0560985.v1\\
	{\em Licensing provisions:} Apache-2.0 \\
	{\em Programming language:}   C++     \\
	{\em Dependencies:}   cmake, Boost, Threading Building Blocks (TBB), SimBody     \\
	{\em Computing platforms:}   Linux, Mac OS, Microsoft Windows     \\
	{\em Support email:}   c.zhang@tum.de, xiangyu.hu@tum.de  \\
\end{small}
%
%
\section{Introduction}\label{sec:introduction}
As an open-source multi-resolution and multi-physics library based on the smoothed particle hydrodynamics (SPH) method, 
SPHinXsys has been developed for modeling the industrial complex systems in the following fields: 
fluid dynamics, 
solid mechanics, 
fluid-structure interaction (FSI), 
thermal and mass diffusion, 
reaction diffusion, 
electromechanics and the beyond. 

The main numerical development features of SPHinXsys v0.2.0 are listed below:
\begin{itemize}
	\item low-dissipation SPH method base on Riemann solvers for violent free-surface flow involving breaking and impact \cite{zhang2017weakly};
	\item dual-criteria time stepping method for weakly compressilbe SPH (WCSPH) method\cite{zhang2020dual};
	\item multi-resolution SPH method for fluid-structure interaction (FSI) \cite{zhang2019multi};
	\item position-based Verlet time stepping scheme \cite{zhang2019multi};
	\item schemes for thermal and mass iso- and anisotropic diffusion \cite{zhang2020integrative};
	\item schemes for reaction-diffusion models \cite{zhang2020integrative};
	\item schemes for electromechanics \cite{zhang2020integrative};
\end{itemize}
SPHinXsys is based on SPH, 
for unified modeling of fluid dynamics, thermal and mass diffusion, reaction diffusion 
 and their coupling with solid mechanics. 
SPH method is a fully Lagrangian particle based method, 
in which the continuum media is discretized into Lagrangian particles 
and the mechanics is approximated as the interaction between them using a kernel function, 
usually a Gaussian-like function. 
As a meshless method, 
SPH does not require a mesh to define the neighboring interaction configuration of particles, 
however, the construction and updated configurations according to the distance between particles is essential. 
A remarkable feature of this method is that its computational algorithm involves a large number of common abstractions, i.e. particles, 
which are suitable to inherently cope with many physical systems.
Due to such unique features, SPHinXsys has intrinsic advantages for modeling multi-physics systems. 
The theory and fundamentals of the SPH method are briefly summarized in Section \ref{sec:sph} 

This paper is structured as follows. 
The main numerical schemes of SPHinXsys dealing with fluid dynamics are summarized in Section \ref{sec:fluiddynamics}, 
solid mechanics in Section \ref{sec:soliddynamics}, 
thermal and mass diffusion in Section \ref{sec:diffusion}, 
reaction-diffusion models in Section \ref{sec:ec-reaction}, 
FSI problems in Section \ref{sec:fsi}, 
position-based Verlet time stepping algorithm in Section \ref{sec:verlet} and 
with the electromechanics in Section \ref{sec:electrical-feedback} .
Furthermore, the source code and the executables are presented in Section \ref{sec:code}. 
The code validations and applications are next summarized in Section \ref{sec:validation} and 
finally, the concluding remarks and future works are noted in Section \ref{sec:conclusion}.  
\subsection{Theory and fundamentals of SPH}\label{sec:sph}
In the SPH method, 
a variable field $f(\mathbf{r})$ in a continuum medium 
is discretized as a particle system   
\begin{equation}
f_i  = \int f(\br) W(\br_i - \br, h)d \br. 
\label{particle-average}
\end{equation}
Here, $i$ is the particle index, $f_i$ the discretized particle-average variable and
$\br_{i}$ the particle position.
The compact-support kernel function $W(\br_{i} - \br, h)$, 
where $h$ is the smoothing length, 
is radially symmetric with respect to $\br_{i}$. 
Since we assume that the mass of each particle $m_i$ is known and invariant (indicating mass conservation),
one has the particle volume $V_i = m_i/\rho_i$, where $\rho_i$ is the particle-average density.

In SPH, 
a variable can be approximated by the particle-average values
\begin{equation}
f(\br) \approx \sum_{j}  V_j W(\br - \br_{j}, h) f_j = \sum_{j}  \frac{m_j}{\rho_j} W(\br - \br_{j}, h) , 
\label{particle-reconstuction}
\end{equation}
where the summation is over all the neighboring particles $j$ located in the support domain of the particle of interest$i$, 
and leads to an approximation of the particle-average density 
\begin{equation}
\rho_i \approx \sum_{j}  m_j W(\mathbf{r}_{i} - \mathbf{r}_{j}, h) = \sum_{j}  m_j W_{ij}, 
\label{particle-density-reconstuction}
\end{equation}
which is an alternative way to the continuity equation for updating the fluid density.

The dynamics of other particle-average variables is based on a general form of interaction between a particle and its neighbors,
i.e. the approximation of the spatial derivative operators on the right hand sides (RHS) of the evolution equations.
The original SPH approximation of the derivative of a variable field $f(\br)$ at particle $i$  is obtained by 
the following formulation
\begin{equation}
\label{eq:gradsph}
\begin{split}
\nabla f_i & \approx \int_{\Omega} \nabla f (\br) W(\br_i - \br, h) dV  \\
& =  - \int_{\Omega} f (\br) \nabla W(\br_i - \br, h) dV \approx  - \sum_{j}  V_j \nabla_i W_{ij}f_j . 
\end{split}
\end{equation}
Here, $\nabla_i W_{ij} = \mathbf{e}_{ij} \frac{\partial W_{ij}}{\partial r_{ij}}$ with $r_{ij}$ and $\mathbf{e}_{ij}$ 
are the distance and unit vector of the particle pair $(i,j)$, respectively.

Note that, Eq. (\ref{eq:gradsph}) can be modified into a strong form
\begin{equation}
\label{eq:gradsph-strong}
\nabla f_i = f_{i}\nabla 1 + \nabla f_i \approx   \sum_{j} V_j \nabla_i W_{ij} f_{ij}, 
\end{equation}
where $f_{ij} = f_{i} - f_{j}$ is the inter-particle difference value. 
The strong-form approximation of the derivative is used to determine the local structure of a field.
Also, with a slight different modification, Eq. (\ref{eq:gradsph}) can be rewritten into a weak form as
\begin{equation}
\label{eq:gradsph-weak}
\nabla f_i = \nabla f_i - f_{i}\nabla 1 \approx   -2\sum_{j}  V_j\nabla W_{ij} \widetilde{f}_{ij}, 
\end{equation}
where $\widetilde{f}_{ij} = \left(f_{i} + f_{j}\right)/2$ is the inter-particle average value. 
The weak-form approximation of derivative is used to compute the surface integration respected to a variable 
for solving its conservation law.
Due to the anti-symmetric property of the derivative of the kernel function, 
i.e. $\nabla_i W_{ij} = - \nabla _j W_{ji}$, it implies the momentum conservation of the particle system.

Since its invention by Lucy \cite{lucy1977numerical} and Gingold and Monaghan \cite{gingold1977smoothed} for modeling astrophysics problems, 
SPH has been successfully exploited in a broad variety of problems 
ranging from fluid dynamics \cite{monaghan1994simulating,hu2006multi,shao2006simulation,zhang2019weakly} 
to solid mechanics \cite{libersky1991smooth,benz1995simulations,monaghan2000sph,randles1996smoothed}, 
fluid-structure interactions (FSI) \cite{antoci2007numerical,  han2018sph, zhang2019multi}, 
and multi-phase flows \cite{rezavand2020weakly}. 
%
%
\section{Schemes for fluid dynamics}\label{sec:fluiddynamics}
\subsection{Governing equation}\label{sec:fluid}
In the Lagrangian frame, the conservation of mass and momentum for fluid dynamics can be written as
\begin{equation} 
\begin{cases}\label{governingeq}
\frac{\text{d} \rho}{\text{d} t}  =  - \rho \nabla \cdot \mathbf{v} \\
\rho \frac{\text{d} \mathbf{v}}{\text{d} t}  =   - \nabla p +  \eta \nabla^2 \mathbf{v} + \rho \mathbf{g}
\end{cases},
\end{equation}
where $\mathbf{v}$ is the velocity, $\rho$ the density, 
$p$ the pressure, $\eta$ the dynamic viscosity, $\mathbf{g}$ the gravity  
and $\frac{\text{d}}{\text{d} t}=\frac{\partial}{\partial t} + \mathbf{v} \cdot \nabla$ stands for material derivative.
For modeling incompressible flow with weakly compressible assumption \cite{monaghan1994simulating,morris1997modeling}, 
an artificial isothermal equation of state (EoS) is introduced to close Eq. (\ref{governingeq})
\begin{equation} \label{eqeos}
p = c^2(\rho - \rho^0).
\end{equation}
With the weakly compressible assumption, the density varies around $1 \% $ \cite{morris1997modeling} 
if an artificial sound speed of $ c = 10 U_{max}$ is employed, 
with $U_{max}$ being the maximum anticipated flow speed.
\subsection{WCSPH method based on Riemann solvers}\label{sec:wcsph}
In SPHinXsys, 
the WCSPH method based on Riemann solvers is applied for fluid dynamics, 
where the continuity and momentum equations are discretized as \cite{zhang2017weakly, zhang2020dual}
\begin{equation} 
	\begin{cases}\label{riemannsph}
		\frac{\text{d} \rho_i}{\text{d} t} = 2\rho_i \sum_j\frac{m_j}{\rho_j}(U^{\ast} - \mathbf{v}_{i}\mathbf{e}_{ij} ) \frac{\partial W_{ij}}{\partial r_{ij}} \\
		\frac{\text{d} \mathbf{v}_i}{\text{d} t}  = - 2\sum_j  m_j\frac{P^{\ast}}{\rho_i \rho_j}  \nabla_i W_{ij}
	\end{cases}. 
\end{equation}
Here, 
$U^{\ast}$ and $P^{\ast}$ are the solutions of inter-particle Riemann problem 
along the unit vector $\mathbf{e}_{ij} = -\mathbf{r}_{ij}/r_{ij}$ pointing from particle $i$ to $j$. 
SPHinXsys applies a low-dissipation Riemann solver \cite{zhang2017weakly}.
	
For viscous flows, the physical shear term can be discretized as \cite{hu2006multi}
\begin{equation}\label{eqviscous}
\bigg( \frac{\text{d} \mathbf{v}_i}{\text{d} t} \bigg)^{(\nu)} 
= 2\sum_j m_j \frac{\eta}{\rho_i \rho_j} \frac{{{\mathbf{v}}_{ij}}}{{{r}_{ij}}} \frac{\partial {{W}_{ij}}}{\partial {{r}_{ij}}}, 
\end{equation}
where $\eta$ is the dynamic viscosity. 

For high Reynolds number flows, the WCSPH method may suffer from tensile instability which induces 
particle clumping or void region.
To remedy this issue, we apply the transport velocity formulation \cite{Adami2013,zhang2017generalized}
\begin{equation}\label{transport}
\frac{\text{d} \mathbf{\overline{v}}_i}{\text{d} t} 
= \frac{\text{d} \mathbf{v}_i}{\text{d} t} - 2\sum_j  m_j\frac{p^0}{\rho_i \rho_j}  \nabla_i W_{ij},
\end{equation}
for modeling flows without free surface.
Here, $p^0$ is the background pressure and $\mathbf{\overline{v}}$ represents the particle transport velocity. 

In SPHinXsys, 
a density initialization scheme is introduced to stabilizes the density 
which is updated by continuity equation in Eq. \ref{riemannsph}.
At the beginning of the new step, the fluid density field of free-surface flows is reinitialized by
\begin{equation} \label{eqrhosum}
	\rho_i = \max(\rho^*, \rho^0 \frac{ \sum W_{ij}}{\sum W^0_{ij}}) ,
\end{equation}
where $\rho^*$ denotes the density before re-initialization and superscript $0$ represents the initial reference value.
For flows without free surface, Eq.  \ref{eqrhosum} is merely modified as 
\begin{equation} \label{eqrhosumnosurface}
	\rho_i =  \rho^0 \frac{ \sum W_{ij}}{\sum W^0_{ij}} .
\end{equation}
\subsection{Wall boundary condition}\label{sec:bc}
In the WCSPH method explained in \ref{sec:wcsph}, 
the interaction between fluid particles and wall particles is determined by
solving a one-sided Riemann problem \cite{zhang2017weakly} along the wall normal direction.

In the one-sided Riemann problem the left state is defined  from the fluid particle corresponding to the local boundary normal,  
\begin{equation}\label{eqriepartial}
	(\rho_L, U_L, P_L)  = (\rho_f,- \mathbf{n}_w \cdot \mathbf{v}_{f},P_f)
\end{equation}
where the subscript $f$ represents the fluid particles and $\mathbf{n}_w$ the local wall normal direction.
According to the physical wall boundary condition the right state velocity $U_R$ is assumed as
\begin{equation}\label{eqrienoslip}
	U_R =  -U_L + 2u_{w},    
\end{equation}
where $u_w$ is the wall velocity. 
Similar to Adami et al. \cite{adami2012generalized} the right state pressure is assumed as   
\begin{equation}\label{eqriepright}
	P_R =  P_L + \rho_{f} \mathbf{g} \cdot \mathbf{r}_{fw}, 
\end{equation}
where $\mathbf{r}_{fw} = \mathbf{r}_w - \mathbf{r}_f$,
and the right state density is obtained by applying the artificial equation of state. 
\subsection{Dual-criteria time stepping}\label{sec:dual-criteria}
SPHinXsys applies the dual-criteria time stepping for the integration of fluid equation. 
Following Ref. \cite{zhang2020dual},  
two time-step size criteria, viz. 
the advection criterion termed $\Delta t_{ad}$ is defined as 
\begin{equation}\label{dt-advection}
	\Delta t_{ad}   =  {CFL}_{ad} \min\left(\frac{h}{|\mathbf{v}|_{max}}, \frac{h^2}{\nu}\right),
\end{equation}
and the acoustic criterion termed $\Delta t_{ac}$ is given by
\begin{equation}\label{dt-relax}
	\Delta t_{ac}   = {CFL}_{ac} \frac{h}{c + |\mathbf{v}|_{max}}. 
\end{equation}
Here, $CFL_{ad} = 0.25$, ${CFL}_{ac} = 0.6$, and
$|\mathbf{v}|_{max}$ is the maximum particle advection velocity in the flow 
while $\nu$ denotes the kinematic viscosity. 
Accordingly, 
the advection criterion controls the updating frequency of particle configuration and 
the acoustic criterion determines the frequency of the pressure relaxation process.
%
%
\section{Schemes for solid dynamics}\label{sec:soliddynamics}
\subsection{Kinematics and governing equation}\label{sec:solid}
The kinematics of the finite deformations can be characterized by introducing a deformation map $\varphi$, 
which maps a material point $\mathbf{\mathbf{r}^0}$ from the initial reference configuration $\Omega^0 \subset \mathbb{R}^d $ 
to the point $\mathbf{r} = \mathbf{\varphi}\left(\mathbf{r}^0, t\right)$ 
in the deformed configuration $\Omega = \mathbf{\varphi} \left(\Omega^0\right)$. 
Here, the superscript $\left( {\bullet} \right)^0$ denotes the quantities in the initial reference configuration. 
Then, the deformation tensor $\mathbb{F}$ can be defined by its derivative with respect to the initial reference configuration as 
\begin{equation} \label{eq:deformationtensor}
\mathbb{F} = \nabla^{0} {\varphi} =  \frac{\partial \varphi}{\partial \mathbf{r}^0}  = \frac{\partial \mathbf{r}}{\partial \mathbf{r}^0} .
\end{equation}
Also, the deformation tensor $\mathbb{F}$ can be calculated from the displacement $\mathbf{u} = \mathbf{r} - \mathbf{r}^0$ through
\begin{equation} \label{eq:deformationtensor-displacement}
\mathbb{F} = \nabla^{0} {\mathbf{u}}  + \mathbb{I},
\end{equation}
where $\mathbb{I}$ represents the unit matrix. 
For incompressible material, we have the constraint
\begin{equation} \label{eq:incompressible}
J = \text{det}\left(\mathbb{F}\right) \equiv 1. 
\end{equation}
Associated with $\mathbb{F}$ are the right and left Cauchy-Green deformation tensors defined by 
\begin{equation} \label{eq:cauchy-green}
\mathbb{C} = \mathbb{F}^{T} \cdot \mathbb{F}\quad \text{and}\quad \mathbb{B} = \mathbb{F} \cdot \mathbb{F}^{T}, 
\end{equation}
respectively. 
Then, four typical invariants of $\mathbb{C}$ (and also of $\mathbb{B}$) can be defined as
\begin{align}\label{eq:invariants}
\mathit{I}_{I} &= \tr \left( \mathbb{C} \right) , & \mathit{I}_{ff} &= \mathbf{f}^0 \cdot \left( \mathbb{C} \mathbf{f}^0 \right), \nonumber \\
\mathit{I}_{ss} &= \mathbf{s}^0 \cdot \left( \mathbb{C} \mathbf{s}^0 \right), & \mathit{I}_{fs} &= \mathbf{f}^0 \cdot \left( \mathbb{C} \mathbf{s}^0 \right) ,
\end{align}
where $\mathbf{f}^0$ and $\mathbf{s}^0$ are the non-deformed myocardial fiber and sheet unit direction, respectively. 
Here, $\mathit{I}_{I} $ is the first principal invariant, 
$\mathit{I}_{ff} $ and $\mathit{I}_{ss} $ are the structure based invariants and $\mathit{I}_{fs}$ is the fiber-sheet shear \cite{holzapfel2009constitutive}.

In a Lagrangian framework, 
the conservation of mass and the linear momentum corresponding to the cardiac mechanics can be expressed as
\begin{equation}\label{eq:mechanical-mom}
\begin{cases}
\rho =  {\rho_0} \frac{1}{J} \quad \\
\rho^0 \frac{\text{d} \mathbf{v}}{\text{d} t}  =  \nabla^{0} \cdot \mathbb{P}^T  + \rho^0 \mathbf{g} \quad  
\end{cases} \Omega^0 \times \left[0, T \right],
\end{equation}
where $\rho$ is the density and $\mathbb{P}$ the first Piola-Kirchhoff stress tensor 
and $\mathbb{P} =  \mathbb{F} \mathbb{S}$ with $\mathbb{S}$ denoting the second Piola-Kirchhoff stress tensor. 
In particular, when the material is linear elastic and isotropic, the constitutive equation is simply given by
\begin{eqnarray}\label{isotropic-linear-elasticity}
	\mathbb{S} & = & K \tr\left(\mathbb{E}\right)  \mathbb{I} + 2 G \left(\mathbb{E} - \frac{1}{3}\tr\left(\mathbb{E}\right)  \mathbb{I} \right) \nonumber \\
	& = & \lambda \tr\left(\mathbb{E}\right) \mathbb{I} + 2 \mu \mathbb{E} ,
\end{eqnarray}
where $\lambda$ and $\mu$ are Lam$\acute{e}$ parameters, 
$K = \lambda + (2\mu/3)$ the bulk modulus and $G = \mu$ the shear modulus. 
The relation between the two modulus is given by
\begin{equation}\label{relation-modulus}
	E = 2G \left(1+2\nu\right) = 3K\left(1 - 2\nu\right)
\end{equation}
with $E$ denoting the Young's modulus and $\nu$ the Poisson ratio. 
Note that the sound speed of solid structure is defined as $c^{S} = \sqrt{K/\rho}$. 
The Neo-Hookean material model can be defined in general form by the strain-energy density function
\begin{eqnarray}\label{Neo-Hookean-energy}
	W  =  \mu \tr \left(\mathbb{E}\right) - \mu \ln J + \frac{\lambda}{2}(\ln J)^{2} .
\end{eqnarray}
Note that the second Piola-Kirchhoff stress $\mathbb{S}$ can be derived as 
\begin{equation}\label{2rd-PK}
	\mathbb{S} = \frac{\partial W}{\partial \mathbb{E}}
\end{equation}
from strain-energy density function. 
\subsection{Total Lagrangian formulation}\label{sec:tlsph}
For solid mechanics, 
SPHinXsys applies the total Lagrangian formulation 
where the initial reference configuration is used for finding the neighboring particles and 
the set of neighboring particles is not altered.  

Firstly, 
a correction matrix $\mathbb{B}^0$ \cite{vignjevic2006sph} is introduced as
\begin{equation} \label{eq:sph-correctmatrix}
\mathbb{B}^0_i = \left( \sum_j V_j \left( \mathbf{r}^0_j - \mathbf{r}^0_i \right) \otimes \nabla^0_i W_{ij} \right) ^{-1} ,
\end{equation}
where 
\begin{equation}\label{strongkernel}
\nabla^0_i W_{ij} = \frac{\partial W\left( |\mathbf{r}^0_{ij}|, h \right)}  {\partial |\mathbf{r}^0_{ij}|} \mathbf{e}^0_{ij}
\end{equation}
denotes the gradient of the kernel function evaluated at the initial reference configuration. 
It is worth noting that the correction matrix is computed in the initial configuration and therefore, 
it is calculated only once before the simulation. 
Then, the momentum conservation equation, Eq.\eqref {eq:mechanical-mom}, can be discretized as 
\begin{equation}\label{eq:sph-mechanical-mom}
\begin{cases}
	\rho_i =  {\rho_0} \frac{1}{\text{det}\left(\mathbb{F}\right) } \quad \\
\frac{\text{d}\mathbf{v}_i}{\text{d}t} = \frac{2}{m_i} \sum_j V_i V_j \tilde{\mathbb{P}}_{ij} \nabla^0_i W_{ij} + \mathbf{g}
\end{cases}, 
\end{equation} 
where the inter-particle averaged first Piola-Kirchhoff stress $\tilde{\mathbb{P}}$ is defined as
\begin{equation}
\tilde{\mathbb{P}}_{ij} = \frac{1}{2} \left( \mathbb{P}_i \mathbb{B}^0_i + \mathbb{P}_j \mathbb{B}^0_j \right). 
\end{equation}
Note that the first Piola-Kirchhoff stress tensor is computed from the constitutive law with the deformation tensor $\mathbb{F}$ is given by
\begin{equation}
\mathbb{F} = \left( \sum_j V_j \left( \mathbf{u}_j - \mathbf{u}_i \right) \otimes \nabla^0_i W_{ij}  \right) \mathbb{B}^0_i + \mathbb{I} .
\end{equation}
%
%
%
\section{Schemes for thermal and mass diffusion}\label{sec:diffusion}
\subsection{Governing equation}
Thermal or mass diffusion, 
in particular anisotropic diffusion, 
occurs in many physical applications, e.g.
thermal conduction in fusion plasma, 
image processing, 
biological processes and medical imaging.
The governing equations for thermal or mass diffusion reads
\begin{equation}\label{eq:diffusion-eq}
\frac{\text{d}C}{\text{d}t}  = \nabla \cdot \left( \mathbb{D} \nabla C \right) , 
\end{equation}
where $C$ is the concentration of a compound and $\mathbb{D}$ is the diffusion coefficient in iso- and aniso-tropic forms. 
\subsection{SPH discretization for the anisotropic diffusion equation}
In SPHinXsys, 
the diffusion equation is discretized by an anisotropic SPH dicretization scheme modified from the work of Tran-Duc et al. \cite{tran2016simulation}. 
Following Ref. \cite{tran2016simulation},  
the diffusion tensor $\mathbb{D}$ is considered to be a symmetric positive-definite matrix and can be decomposed by Cholesky decomposition as 
\begin{equation}\label{eq:chol}
\mathbb{D} = \mathbb{L} \mathbb{L}^T
\end{equation}
where $\mathbb{L}$ is a lower triangular matrix with real and positive diagonal entries and $\mathbb{L}^T$ denotes the transpose of $\mathbb{L}$. 
Eq. \eqref{eq:diffusion} can be rewritten in isotropic form as 
\begin{equation}\label{eq:diffusion-trans}
\nabla \cdot (\mathbb{D} \nabla) =  \nabla \cdot (\mathbb{\mathbb{L} \mathbb{L}^T} \nabla) 
=  (\mathbb{L}^T \nabla)^T \cdot (\mathbb{L}^T\nabla) = \widetilde{\nabla}^2, 
\end{equation}
where $ \widetilde{\nabla} = \mathbb{L}^T \nabla$.
Then, the new isotropic diffusion operator is approximated by the following kernel integral by neglecting the high-order term
\begin{equation}\label{eq:diffusion-int}
\widetilde{\nabla} \cdot (\widetilde{\nabla})  C =   2 \int_{\Omega} \frac{C (\widetilde{\mathbf{r}}) - C (\widetilde{\mathbf{r}'}) }{|\widetilde{\mathbf{r}} - \widetilde{\mathbf{r}'}|} \frac{\partial W\left( \widetilde{\mathbf{r}} - \widetilde{\mathbf{r}'}, \widetilde{h}\right) }{\partial | \widetilde{\mathbf{r}} - \widetilde{\mathbf{r}'} | } d \widetilde{\mathbf{r}}, 
\end{equation}
where $\widetilde{\mathbf{r}} = \mathbb{L}^{-1} \mathbf{r}$ and $\widetilde{h} = \mathbb{L}^{-1} h$ . 
Upon the coordinate transformation, 
the kernel gradient can be rewritten as
\begin{equation}\label{eq:diffusion-kernel-trans}
\frac{\partial W\left( \widetilde{\mathbf{r}} - \widetilde{\mathbf{r}'}, \widetilde{h}\right) }{\partial \left( \widetilde{\mathbf{r}} - \widetilde{\mathbf{r'}}\right) } =
\frac{1}{|\mathbb{L}^{-1}||\mathbb{L}^{-1} \mathbf{e}_{\widetilde{\mathbf{r}\mathbf{r}}|}} \frac{\partial W \left( {\mathbf{r}} - {\mathbf{r'}}\right)}{\partial |{\mathbf{r}} - {\mathbf{r'}}|}
\end{equation}
with $\mathbf{e}_{{\mathbf{r'}} \mathbf{r}} = \frac{{\mathbf{r'}} - \mathbf{r}}{|{\mathbf{r'}} - \mathbf{r}|}$. 
Finally, 
Eq. \eqref{eq:diffusion-trans} can be discretized in SPH form as 
\begin{equation}\label{grad-laplace}
\begin{split}
\widetilde{\nabla}^2 C  & \approx  2 \sum_{j}^{N} \frac{m_j}{\rho_j} \bigg(C(\mathbf{r_i}) - C(\mathbf{r_j}) \bigg) \frac{1}{(\widetilde{\mathbb{L}}_{ij} \mathbf{e}_{ij})^2} \frac{1}{r_{ij}} \frac{\partial W_{ij}}{\partial r_{ij}}
\end{split},
\end{equation}
by replacing the term $\mathbb{L}_{ij}^{-1}$ with its linear approximation 
$\widetilde{\mathbb{L}}_{ij}$ given by 
\begin{equation}
\widetilde{\mathbb{L}}_{ij} = \frac{\widetilde{\mathbb{L}}_{i}\widetilde{\mathbb{L}}_{j} }{\widetilde{\mathbb{L}}_{i} + \widetilde{\mathbb{L}}_{j} }
\end{equation}
where $\widetilde{\mathbb{L}}_{i}$ is defined as 
\begin{equation}
\widetilde{\mathbb{D}}_{i} = \left( \widetilde{\mathbb{L}}^{-1}_{i}\right)  \left( \widetilde{\mathbb{L}}^{-1}_{i}\right) ^T .
\end{equation}
In this case, 
the Cholesky decomposition and the corresponding matrix inverse are computed once for each particle before the simulation. 
Also, 
Eq. \eqref{grad-laplace} can be rewritten by introducing a kernel correction matrix Eq. \eqref{eq:sph-correctmatrix} as
\begin{equation}\label{grad-laplace-renormalize}
	\frac{\text{d}C}{\text{d}t}   = 2 \sum_{j} \frac{m_j}{\rho_j} \bigg(\mathbb{B}^0_i C(\mathbf{r_i}) - \mathbb{B}^0_j C (\mathbf{r_j}) \bigg) \cdot \frac{\mathbf{e}_{ij} \cdot \mathbf{e}_{ij}}{(\widetilde{\mathbb{L}}_{ij} \mathbf{e}_{ij})^2} \frac{1}{r_{ij}} \frac{\partial W_{ij}}{\partial r_{ij}} .
\end{equation}
%
%
\section{Schemes for reaction-diffusion model}\label{sec:ec-reaction}
\subsection{Governing equation}\label{sec:ecr}
In recent years, 
the reaction-diffusion model has attracted a considerable deal of attention due to its ubiquitous application in many fields of science. 
The reaction-diffusion model can generate a wide variety of spatial patterns, 
which has been widely applied in chemistry, biology, and physics, 
even used to explain self-regulated pattern formation in the developing animal embryo. 
The general form of reaction-diffusion model reads 
\begin{equation}\label{eq:diffusion-reaction}
\frac{\text{d}V}{\text{d}t}  = \nabla \cdot (\mathbb{D} \nabla V) + I(V) , 
\end{equation}
where $\mathbb{D}$ is diffusion tensor and $I(V) $ a nonlinear function. 
For $I(V)  = V - V^3$, 
Eq. \eqref{eq:diffusion-reaction} becomes the Allen–Cahn equation, which describes the mixture of two incompressible fluids. 
When the FitzHugh-Nagumo model \cite{fitzhugh1961impulses} is applied, 
Eq. \eqref{eq:diffusion-reaction} becomes the well known mono domain equation \cite{quarteroni2017cardiovascular} which describes the cell electrophysiological dynamics. 
Electrophysiological dynamics of the heart describe how electrical currents flow through the heart, 
controlling its contractions, and are used to ascertain the effects of certain drugs designed to treat, for example, arrhythmia. 

The Fitzhugh-Nagumo (FN) model reads \cite{fitzhugh1961impulses} 
\begin{equation}\label{eq:fhn}
\begin{cases}
I(V, w) = -V(V - a)(V - 1) - w \\
\dot{w} = g(V, w) = \epsilon_0 ( \beta V - \gamma w - \sigma)
\end{cases},
\end{equation}
where $\epsilon_0$, $\beta$, $\gamma$ and $\sigma$ are suitable constant parameters, given specifically. 

The Aliev-Panfilow (AP) model \cite{aliev1996simple}, 
a variant of the FN model, 
has been successfully implemented in the simulations of ventricular fibrillation in real heat geometries \cite{panfilov1999three}
and it is particularly suitable for the problems where electrical activity of the heart is of the main interest. 
The AP model has the following form
\begin{equation}\label{eq:a-p}
\begin{cases}
I(V, w) = -k V(V - a)(V - 1) - w V \\
\dot{w} = g(V, w) = \epsilon(V, w)(-k V (V - b - 1) - w)
\end{cases},
\end{equation}
where $\epsilon(V_m, w) = \epsilon_0 + \mu_1 w / (\mu_2 + V_m)$ and $k$, $a $, $b$,  $\epsilon_0$, $\mu_1$ and $\mu_2$ are suitable constant parameters. 
\subsection{SPH method for reaction-diffusion}\label{sec:sph-ecr}
The reaction-diffusion model consists of a coupled system of partial differential equations (PDE) governing the diffusion process 
as well as ordinary differential equations (ODE) governing the reactive kinetics of the gating variable. 
In SPHinXsys, 
the operator splitting method \cite{quarteroni2017cardiovascular} is applied and results in a PDE governing the anisotropic diffusion
\begin{equation}\label{eq:diffusion}
\frac{\text{d}V}{\text{d}t}  = \nabla \cdot (\mathbb{D} \nabla V), 
\end{equation}
and two ODEs 
\begin{equation}\label{eq:ode-system}
\begin{cases}
\frac{\text{d}V}{\text{d}t}  = I(V, w) \\
\frac{\text{d}w}{\text{d}t} = g(V, w) 
\end{cases},  
\end{equation}
where $I(V, w)$ and $ g(V, w)$ are defined by the FN model Eq. \eqref{eq:fhn} or the AP model Eq. \eqref{eq:a-p}. 
The schemes for discretizing the diffusion equation is presented in Section \ref{sec:diffusion}.
\subsection{Reaction-by-reaction splitting}
In SPHinXsys, a reaction-by-reaction splitting method \cite{wang2019split} is introduced for solving 
the system of ODEs defined by Eq. \eqref{eq:ode-system}, which is generally stiff and induces numerical instability when the integration time step is not sufficiently small.  
The multi-reaction system can be decoupled using the second-order accurate Strange splitting as
\begin{equation}\label{eq:ode-spliting-2rd}
R^{(\Delta t)} = R_V^{(\frac{\Delta t}{2})} \circ R_w^{(\frac{\Delta t}{2})} \circ R_w^{(\frac{\Delta t}{2})} \circ R_V^{(\frac{\Delta t}{2})},
\end{equation}
where the $ \circ $ symbol separates each reaction and 
indicates that the operator $R_V^{(\Delta t)}$ is applied after $R_w^{(\Delta t)}$. 
Note that the reaction-by-reaction splitting methodology can be extended to more complex ionic models, 
e.g. the Tusscher-Panfilov model \cite{ten2004model}. 

Following Ref. \cite{wang2019split}, we rewrite the Eq. \eqref{eq:ode-system} in the following form 
\begin{equation}\label{eq:ode-new-form}
\frac{\text{d} y}{\text{d} t} = q(y,t) - p(y,t) y, 
\end{equation}
where $q(y,t)$ is the production rate and $ p(y,t) y$ is the loss rate \cite{wang2019split}.
The general form of Eq. \eqref{eq:ode-new-form}, where the analytical solution is not explicitly known or difficult to derive, 
can be solved by using the quasi steady state (QSS) method for an approximate solution as
\begin{equation}\label{eq:ode-qss}
y^{n + 1} = y^n e^{-p(y^n, t) \Delta t } + \frac{q(y^n, t) }{p(y^n, t)} \left(1 - e^{-p(y^n, t) \Delta t} \right).
\end{equation}
Note that the QSS method is unconditionally stable due to the analytic form, 
and thus a larger time step is allowed for the splitting method, 
leading to a higher computational efficiency. 
%
%
\section{Multi-resolution FSI coupling} \label{sec:fsi}
SPHinXsys benefits from a multi-resolution framework, i.e. 
the fluid and solid equations are discretized by different spatial-temporal resolutions, 
for modeling FSI problems. 
More precisely, 
different particle spacing, hence different smoothing lengths, 
and different time steps are utilized to discretize the fluid and solid equations \cite{zhang2019multi}. 

In the multi-resolution framework, the governing equations of 
Eq. \eqref{governingeq} are discretized as
\begin{equation} 
	\begin{cases}\label{eq:mr-fluid}
		\frac{\text{d} \rho_i}{\text{d} t} = 2\rho_i \sum_j\frac{m_j}{\rho_j}(U^{\ast} - \mathbf{v}_{i}\mathbf{e}_{ij} ) \frac{\partial W_{ij}^{h^F}}{\partial r_{ij}} \\
		\frac{\text{d} \mathbf{v}_i}{\text{d} t}  = - 2\sum_j  m_j\frac{P^{\ast}}{\rho_i \rho_j}  \nabla_i W_{ij}^{h^F}  + \mathbf{g} + \mathbf{f}_i^{S:p}\left(h^F\right) + \mathbf{f}_i^{S:v}\left(h^F\right)
	\end{cases},
\end{equation}
where $h^F$ represents the smoothing length used for fluid. 

Also, the discretization of solid equation of Eq. \eqref{eq:sph-mechanical-mom} is modified to
\begin{equation}\label{eq:mr-solid}
	\begin{cases}
		\rho_a =  {\rho_0} \frac{1}{\text{det}\left(\mathbb{F}\right) } \quad \\
		\frac{\text{d}\mathbf{v}_a}{\text{d}t} = \frac{2}{m_a} \sum_b V_a V_b \tilde{\mathbb{P}}_{ab} \nabla^0_a W_{ab} + \mathbf{g} + \mathbf{f}_a^{F:p}\left(h^F\right) + \mathbf{f}_a^{F:v}\left(h^F\right)
	\end{cases},
\end{equation} 
where $h^S$ denotes the smoothing length used for solid discretization. 
Generally,
we use $h^F$ assuming $h^F \geqslant h^S$. 
In more details, 
the forces $\mathbf{f}_i^{S:p}\left(h^F\right)$ and $\mathbf{f}_i^{S:v}\left(h^F\right)$ of Eq. \eqref{eq:mr-fluid} are modified to
\begin{equation}\label{fs-force-mr-sp}
\mathbf{f}_i^{S:p}\left(h^F\right) =  - 2  \sum_a V_i V_a \frac{p_i \rho_a^d + p^d_a \rho_i}{\rho_i + \rho^d_a} \nabla_i W(\mathbf{r}_{ia}, h^F )
\end{equation}
and
\begin{equation}\label{fs-force-mr-sv}
\mathbf{f}_i^{S:v}\left(h^F\right)= 2\sum_a  \eta V_i V_a \frac{\mathbf{v}_i - \mathbf{v}^d_a}{|\mathbf{r}_{ia}| + 0.01h} \frac{\partial W(\mathbf{r}_{ia}, h^F )}{\partial {{r}_{ia}}}.
\end{equation}
The fluid forces exerting on the solid structure $\mathbf{f}_a^{F:p}\left(h^F\right) $ and $\mathbf{f}_a^{F:v}\left(h^F\right)$ are straightforward to derive. 

The multi time stepping scheme for FSI is coupled with the dual-criteria time stepping presented in Section \ref{sec:dual-criteria}. 
More precisely, 
during each fluid acoustic time step (Eq. \eqref{dt-relax}) 
the structure time integration $\kappa = [\frac{\Delta t_{ac}^F}{\Delta t^S}] + 1$ marches with the
solid time-step criterion
\begin{equation}\label{dts-advection}
\Delta t^S   =  0.6 \min\left(\frac{h^S}{c^S + |\mathbf{v}|_{max}},
\sqrt{\frac{h^S}{|\frac{\text{d}\mathbf{v}}{\text{d}t}|_{max}}} \right).
\end{equation}
As different time steps are applied in the integration of fluid and solid equations, 
we redefine the imaginary pressure $p_a^d$ and velocity $\mathbf{v}_a^d$ in Eqs. \eqref{fs-force-mr-sp} and \eqref{fs-force-mr-sv} as
\begin{equation} \label{fs-coupling-mr }
\begin{cases}
p_a^d = p_i + \rho_i max(0, (\mathbf{g} - \widetilde{\frac{\text{d} \mathbf{v}_a}{\text{d}t}}) \cdot \mathbf{n}^S) (\mathbf{r}_{ia} \cdot \mathbf{n}^S) \\
\mathbf{v}_a^d = 2 \mathbf{v}_i  - \widetilde{\mathbf{v}}_a
\end{cases}, 
\end{equation}
where \eqref{fs-coupling-mr } $\widetilde{\mathbf{v}}_a$ and $\widetilde{\frac{d\mathbf{v}_a}{dt}}$ 
represent the single averaged velocity and acceleration of solid particles during a fluid acoustic time step. 
%
%
\section{Position-based Verlet scheme} \label{sec:verlet}
SPHinXsys applies the position-based Verlet scheme for the time integration of fluid and solid equation.
As presented in Ref. \cite{zhang2019multi}, 
the position-based Verlet achieves strict momentum conservation in fluid-structure coupling when multiple time steps is employed.
In the position-based Verlet scheme,
a half step for position is followed by a full step for velocity and another half step for position. 
Denoting the values at the beginning of a fluid acoustic time step by superscript $n$, 
at the mid-point by $n + \frac{1}{2}$ and eventually at the end of the time-step by $n + 1$, here we 
summarize the scheme.
At first, the integration of the fluid is conducted as
\begin{equation}\label{verlet-first-half}
	\begin{cases}
		\rho_i^{n + \frac{1}{2}} = \rho_i^n + \frac{1}{2}\Delta t_{ac}^F  \frac{d \rho_i}{dt}\\
		\mathbf{r}_i^{n + \frac{1}{2}} = \mathbf{r}_i^n + \frac{1}{2} \Delta t_{ac}^F {\mathbf{v}_i}^{n}
	\end{cases}, 
\end{equation}
by updating the density and position fields into the mid-point. 
The particle velocity is next updated to the new time step in the following form
\begin{equation}\label{verlet-first-mediate}
	\mathbf{v}_i^{n + 1} = \mathbf{v}_i^{n} +  \Delta t_{ac}^F  \frac{d \mathbf{v}_i}{dt}. 
\end{equation}
Finally, the position and density of fluid particles are updated to the new time step by 
\begin{equation}\label{verlet-first-final}
	\begin{cases}
		\mathbf{r}_i^{n + 1} = \mathbf{r}_i^ {n + \frac{1}{2}} +  \frac{1}{2} \Delta t_{ac}^F {\mathbf{v}_i} \\
		\rho_i^{n + 1} = \rho_i^{n + \frac{1}{2}} + \frac{1}{2} \Delta t_{ac}^F \frac{d \rho_i}{dt}
	\end{cases}. 
\end{equation}

For solid equations, 
index $\varkappa = 0, 1, ...,  \kappa-1 $ is used to denote integration step for solid particles.  
With the position-based Verlet scheme, 
the deformation tensor, density and particle position are updated to the midpoint as 
\begin{equation}\label{verlet-first-half-solid}
	\begin{cases}
		\mathbb{F}_a^{\varkappa + \frac{1}{2}} = \mathbb{F}_a^{\varkappa} + \frac{1}{2} \Delta t^S \frac{\text{d} \mathbb{F}_a}{\text{d}t}\\
		\rho_a^{\varkappa + \frac{1}{2}} = \rho_a^0 \frac{1}{J} \\
		\mathbf{r}_a^{\varkappa + \frac{1}{2}} = \mathbf{r}_a^{\varkappa} + \frac{1}{2} \Delta t^S {\mathbf{v}_a}
	\end{cases}. 
\end{equation}
The velocity is next updated by
\begin{equation}\label{verlet-first-mediate-solid}
	\mathbf{v}_a^{\varkappa + 1} = \mathbf{v}_a^{\varkappa} +  \Delta t^S  \frac{d \mathbf{v}_a}{dt}. 
\end{equation}
Lastly, the deformation tensor and position of solid particles are updated to the new time step of the solid structure with 
\begin{equation}\label{verlet-first-final-solid}
	\begin{cases}
		\mathbb{F}_a^{\varkappa + 1} = \mathbb{F}_a^{\varkappa + \frac{1}{2}} + \frac{1}{2} \Delta t^S \frac{\text{d} \mathbb{F}_a}{\text{d}t}\\
		\rho_a^{\varkappa + 1} = \rho_a^0 \frac{1}{J} \\
		\mathbf{r}_a^{\varkappa + 1} = \mathbf{r}_a^{\varkappa + \frac{1}{2}} + \frac{1}{2} \Delta t^S {\mathbf{v}_a}^{\varkappa + 1}
	\end{cases}. 
\end{equation}
%
%
%
\section{Electromechanics} \label{sec:electrical-feedback}
As the multi-physics modeling of the complete cardiac process, including the muscle tissues,
is one of the main future goals of SPHinXsys,
the active stress approach is applied for modeling the electromechanics.
Following the work of Nash and Panfilov \cite{nash2004electromechanical}, 
the stress tensor is coupled with the transmembrane potential $V_m$ through the active stress approach, 
which decomposes the first Piola-Kirchhoff stress $\mathbb{P}$ into passive and active parts 
\begin{equation}
\mathbb{P} = \mathbb{P}_{passive} + \mathbb{P}_{active}. 
\end{equation}
Here,  
the passive component $\mathbb{P}_{passive}$ describes the stress required to obtain a given deformation of the passive myocardium, 
and an active component $\mathbb{P}_{active}$ denotes the tension generated by the depolarization 
of the propagating transmembrane potential. 

For the passive mechanical response,    
we consider the Holzapfel-Odgen model, which proposed the following strain energy function, considering different contributions and taking the anisotropic nature of the myocardium into account.
To ensure that the stress vanishes in the reference configuration and encompasses the finite extensibility, 
we modify the strain-energy function as 
\begin{eqnarray}\label{eq:new-muscle-energy}
\mathbf{W}  & = &  \frac{a}{2b}\exp\left[b (I_1 - 3 )\right] - a \ln J  + \frac{\lambda}{2}(\ln J)^{2} \nonumber \\
& + & \sum_{i = f,s} \frac{a_i}{2b_i}\{\text{exp}\left[b_i\left(\mathit{I}_{ii}- 1 \right)^2\right] - 1\} \nonumber \\
& + & \frac{a_{fs}}{2b_{fs}}\{\text{exp}\left[b_{fs}\mathit{I}^2_{fs} \right] - 1\} ,
\end{eqnarray}
where $a$, $b$, $a_f$, $b_f$, $a_s$, $b_s$, $a_{fs}$ and $b_{fs}$ are eight positive material constants, 
with the $a$ parameters having the dimension of stress and $b$ parameters being dimensionless. 
Here, the second Piola-Kirchhoff stress $\mathbb{S}$ is defined by
\begin{equation}\label{eq:second-PK}
\begin{split}
\mathbb{S}  = 2 \frac{\partial \mathbf{W}}{\partial \mathbb{C}} -p\mathbb{C}^{-1} & = 2\sum_{j} \frac{\partial \mathbf{W}}{\partial \mathit{I}_j} \frac{\partial \mathit{I}_j}{\partial \mathbb{C}} -p\mathbb{C}^{-1} \\
& \quad j = I,ff,ss,fs,
\end{split}
\end{equation}
where 
\begin{align}\label{eq:second-PK-2}
&\frac{\partial \mathit{I}_1}{\partial \mathbb{C}} =  \mathbb{I},
& \frac{\partial \mathit{I}_{ff}}{\partial \mathbb{C}} = &\mathbf{f}_0 \otimes \mathbf{f}_0,  \\
&\frac{\partial \mathit{I}_{ss}}{\partial \mathbb{C}} =  \mathbf{f}_0 \otimes \mathbf{f}_0,  
 & \frac{\partial \mathit{I}_{fs}}{\partial \mathbb{C}} = & \mathbf{f}_0 \otimes \mathbf{s}_0 + \mathbf{s}_0 \otimes \mathbf{f}_0 ,
\end{align}
and $p$ is the Lagrange multiplier arising from the imposition of incompressibility. 
Substituting Eqs. \eqref{eq:second-PK} and \eqref{eq:second-PK-2} into Eq.\eqref{eq:new-muscle-energy} the second Piola-Kirchhoff stress is given as
\begin{eqnarray}
\mathbb{S} & = & a \text{exp} \left[b\left({\mathit{I}}_{I} - 3 \right)\right] + \left\{ \lambda\ln J - a \right\}\mathbb{C}^{-1} \nonumber\\ 
& + & 2a_f \left({\mathit{I}}_{f}- 1 \right)  \text{exp}\left[b_f\left({\mathit{I}}_{f}- 1 \right)^2\right] \mathbf{f}_0 \otimes \mathbf{f}_0  \nonumber \\
& + & 2a_s \left({\mathit{I}}_{s}- 1 \right)  \text{exp}\left[b_s\left({\mathit{I}}_{s}- 1 \right)^2\right] \mathbf{s}_0 \otimes \mathbf{s}_0  \nonumber \\
& + & a_fs {\mathit{I}}_{fs} \text{exp}\left[b_fs\left({\mathit{I}}_{fs}\right)^2\right] \mathbf{fs}_0 \otimes \mathbf{fs}_0 .
\end{eqnarray}

Following the active stress approach proposed by Nash and Panfilov \cite{nash2004electromechanical}, 
the active component provides the internal active contraction stress by
\begin{equation}
\mathbb{P}_{active} = T_a \mathbb{F} \mathbf{f}_0 \otimes \mathbf{f}_0,
\end{equation}
where $T_a$ represents the active magnitude of the stress and its evolution is given by an ODE as 
\begin{equation}
\dot{T_a} = \epsilon\left(V_m\right)\left[k_a\left(V_m - {V}_r \right) - T_a\right], 
\end{equation}
where parameters $k_a$ and ${V}_r$ control the maximum active force, the resting action potential 
and the activation function \cite{wong2011computational}
\begin{equation}
\epsilon\left(V_m \right) = \epsilon_0 + \left(\epsilon_{\infty} -\epsilon_{-\infty} \right) \text{exp}\{-\text{exp}\left[-\xi\left(V_m - \overline{V}_m\right)\right]\}.
\end{equation} 
Here, the limiting values $\epsilon_{-\infty}$ at $V_m \rightarrow -\infty$ and $\epsilon_{\infty}$ at $V_m \rightarrow \infty$, 
the phase shift $\overline{V}_m$ and the transition slope $\xi$ will ensure a smooth activation of the muscle traction. 
%
%
\section{Source code}\label{sec:code}
In this section, 
an overview of SPHinXsys is given in Section \ref{sec:overview}, 
a synthetic description of the program units is reported in Section \ref{sec:code-unit} 
and a brief summary of the installation instructions is given in Section \ref{sec:code-compilation}. 
For more details, the readers are referred to the documentation presented in SPHinXsys's repository.
\subsection{Overview}\label{sec:overview}
\begin{figure*}[tb!]
	\centering
	\includegraphics[trim = 1cm 4cm 1cm 1.5cm, clip,width=0.85 \textwidth]{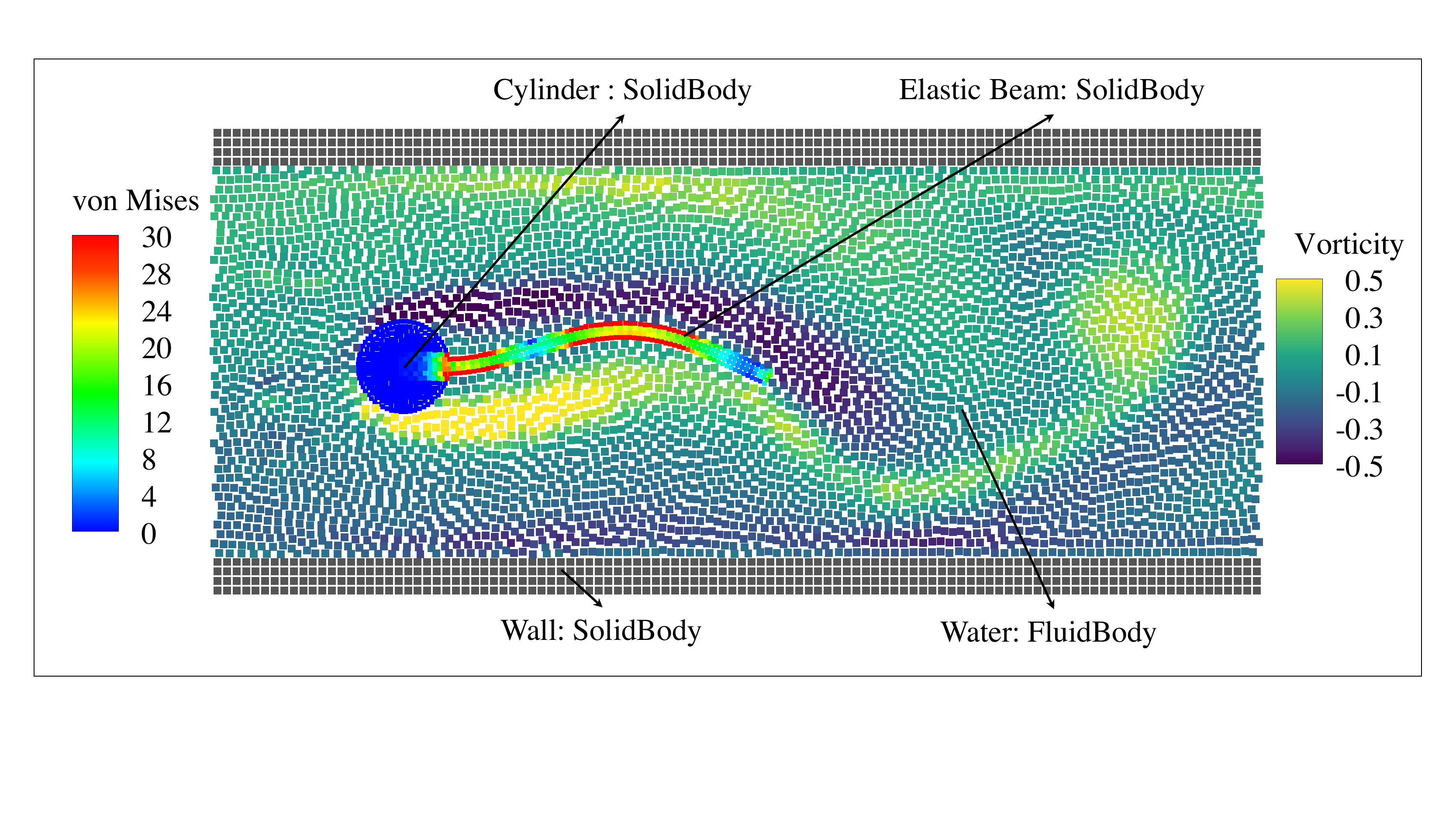}
	\caption{A typical FSI problem involving a rigid solid (wall) body, 
		a composite solid (obstacle) body and a fluid body. The wall body has two (upper and lower) components. 
		The obstacle body is composed of a rigid (cylinder) and an elastic (beam) components.}
	\label{figs:fsi}
\end{figure*}

In SPHinxSys, the whole computational domain is modeled as SPH bodies and each body is composed of an ensemble of SPH particles.
See Figure \ref{figs:fsi} for a typical example in the simulation of flow induced vibration of a flexible beam attached to a rigid cylinder.  
Note that a SPH solid body may be composed of more than one components.
As shown in Fig. \ref{figs:fsi}, 
while the wall body has two rigid solid components,
the solid obstacle is composed of a rigid and an elastic component.
The particle interaction is decomposed into two parts: 
1) inner interaction, which represents the interaction of two neighboring particles located in the same SPH body;
and 2) contact interaction, which denotes the interaction of two neighboring particles originated from different SPH bodies. 
%
\subsection{Synthetic description of the program units}\label{sec:code-unit}
Folders of SPHinXsys repository are reported in Table \ref{tab:folder}. 
The program units of SPHinXsys v0.2.0 (folder "SPHinXsys'') are grouped in subfolders, 
associated with the following topics: 
data structure and general functions for vector and scalar data (Table \ref{tab:data}); 
SPHBody and the derived FluidBody and SolidBody (Table \ref{tab:sphbody}); 
geometry representation of SPHBody (Table \ref{tab:geometry}); 
material property (Table \ref{tab:material}); 
Kernel function (Table \ref{tab:kernel}); 
meshes for level-set and cell-linked lists (Table \ref{tab:mesh}); 
particle container (Table \ref{tab:particle}); 
particle generator (Table \ref{tab:particle-generator}); 
particle dynamics (Table \ref{tab:particle-dynamics}); 
interface for Simbody (Table \ref{tab:simbody}) and in- and output system (``I/O", Table \ref{tab:io}).

The optional input files of SPHinXsys are listed here:
\begin{itemize}
	\item ensemble of points and their volumes for generating particles directly
	\item ensemble of points in 2D for representing polygon
	\item ploymesh in STL or OBJ format for representing 3D geometry
\end{itemize}

The main output files of SPHinXsys report the following information:
\begin{itemize}
	\item basic parameters for test cases
	\item particle data in PLT or VTU formats for different visualization methods
	\item time series of the observed variables for the monitoring probes
\end{itemize}
\begin{table*}[htb!]
	\centering
	\caption{Folders of SPHinXsys v0.2.0 repository. }
	\begin{tabular}{ cc}
		\hline
		Folder   &  Description \\ 
		\hline
		(repository folder) & Readme, installation instruction, Doxyfile and Apache license file \\
		doc						  & Documentation file \\		
		cmake 				   & cmake files for multi-platform compilation and dependencies search\\		
		cases\_test 		& Executable code for test cases\\		
		SPHinxsys 		  & Source code for the CPU compilation \\
		\hline	
	\end{tabular}
	\label{tab:folder}
\end{table*}
\begin{table*}[htb!]
	\centering
	\caption{Program units (".h" and ".cpp") for the basic data types in folder "SPHinxsys/src/share/base\_data\_type". }
	\begin{tabular}{cc}
		\hline
		Program unit  &  Description \\ 
		\hline
		large\_data\_container			& Definition of vector and matrix data structures\\		
		scalar\_functions 				   &Functions for scalar data, e.g., ABS, AMAX, AMIN and SGN \\		
		small\_vectors 		& Functions for vector data, e.g. sorting, shuffle, difference and subtract \\
		\hline	
	\end{tabular}
	\label{tab:data}
\end{table*}
\begin{table*}[htb!]
	\centering
	\caption{Program units (".h" and ".cpp") for the SPHBody in folder "SPHinxsys/src/share/bodies". }
	\begin{tabular}{ cc}
		\hline
		Program unit   &  Description \\ 
		\hline
		base\_body			& Base class of SPHBody \\		
		fluid\_body 		& Fluid-like SPHBody which contains fluid property \\		
		solid\_body 		& Solid-like SPHBody  which contains solid property \\
		body\_relation 	 & Relationship, inner or contact interaction, between bodies \\		
		\hline	
	\end{tabular}
	\label{tab:sphbody}
\end{table*}
\begin{table*}[htb!]
	\centering
	\caption{Program units (".h" and ".cpp") for the geometry representation in folder "SPHinxsys/src/share/base\_geometry". }
	\begin{tabular}{ cc}
		\hline
		Program unit   &  Description \\ 
		\hline
		base\_geometry & Base class of geometry \\
		\hline
		geometry			
		&
		\begin{tabular}{@{}c@{}} 
			Geometry representation of SPHBody \\ 
			2D geometry is based on polygon provided by Boost library \\
			3D geometry is based on parsing polymesh in STL or OBJ format \\
		\end{tabular} \\
		\hline	
	\end{tabular}
	\label{tab:geometry}
\end{table*}
\begin{table*}[htb!]
	\centering
	\caption{Program units (".h" and ".cpp") for the Kernel in folder "SPHinxsys/src/share/kernels". }
	\begin{tabular}{ cc}
		\hline
		Program unit   &  Description \\ 
		\hline
		base\_kernel			      		 & Base class of Kernel function \\		
		kernel\_hyperbolic	    		&  Hyperbolic kernel \cite{yang2014smoothed} \\		
		kernel\_wenland\_c2 		& Wenland C2 kernel \cite{wendland1995piecewise}\\
		\hline	
	\end{tabular}
	\label{tab:material}
\end{table*}
\begin{table*}[htb!]
	\centering
	\caption{Program units (".h" and ".cpp") for material in folder "SPHinxsys/src/share/materials". }
	\begin{tabular}{ cc}
		\hline
		Program unit   &  Description \\ 
		\hline
		base\_material			      		 & Base class of material \\		
		weakly\_compressible\_fluid	    		& Fluid with weakly-compressible assumption \\		
		elastic\_solid 		&Elastic solid\\
		diffusion\_reaction & Reaction-diffusion model\\
		\hline	
	\end{tabular}
	\label{tab:kernel}
\end{table*}
\begin{table*}[htb!]
	\centering
	\caption{Program units (".h" and ".cpp") for the mesh in folder "SPHinxsys/src/share/meshes". }
	\begin{tabular}{ cc}
		\hline
		Program unit   &  Description \\ 
		\hline
		base\_mesh			      		 	& Base class of Mesh and the derived level-set mesh \\		
		mesh\_cell\_linked\_list  &  Background mesh for cell-linked lists \\
		\hline	
	\end{tabular}
	\label{tab:mesh}
\end{table*}
\begin{table*}[htb!]
	\centering
	\caption{Program units (".h" and ".cpp") for the particle in folder "SPHinxsys/src/share/particles". }
	\begin{tabular}{ cc}
		\hline
		Program unit   &  Description \\ 
		\hline
		base\_particles		 & Class of BaseParticles which contains position, velocity and volume data  \\		
		fluid\_particles	  &  Particles which contains fluid properties, e.g. density, mass and pressure \\
		solid\_particles	 &  Particles which contains elastic solid properties, e.g. density, mass and stress \\
		diffusion\_reaction\_particles	 & Particles which contains diffusion-reaction properties, e.g. species\\
		neighbor\_relation	 &  Particle interaction configuration, including kernel value and the gradients\\
		\hline	
	\end{tabular}
	\label{tab:particle}
\end{table*}
\begin{table*}[htb!]
	\centering
	\caption{Program units (".h" and ".cpp") for the particle generator in folder "SPHinxsys/src/share/particle\_generator". }
	\begin{tabular}{ cc}
		\hline
		Program unit   &  Description \\ 
		\hline
		base\_particle\_generator		& 
						\begin{tabular}{@{}c@{}} 
				Base class of particle generator \\		
				Particle direct generator by reading particle position and volume
			\end{tabular} \\
		\hline
		particle\_generator\_lattice  &  Particle generator on Lattice point\\
		\hline	
	\end{tabular}
	\label{tab:particle-generator}
\end{table*}
%
%
\begin{table*}[htb!]
	\centering
	\caption{Program units (".h" , ".hpp" and ".cpp") for the particle dynamics in folder "SPHinxsys/src/share/particle\_dynamics". }
	\begin{tabular}{ cc}
		\hline
		Program unit   &  Description \\ 
		\hline
		base\_particle\_dynamics			  & 
			\begin{tabular}{@{}c@{}} 
				Base classes of particle dynamics describing the particle interaction which \\ 
				defines differential operators for fluxes in continuum mechanics
			\end{tabular} \\		
		\hline
		particle\_dynamics\_algorithms & 
			\begin{tabular}{@{}c@{}} 
				Particle dynamics describe 3 types of particle interaction: \\ 
				1) Simple dynamics for updating particle state without particle-interaction\\
				2) Inner dynamics for  inner particle-interaction \\
				3) Complex dynamics for  inner and contact particle-interactions
			\end{tabular}  \\				
		\hline			
		\begin{tabular}{@{}c@{}} 
			external\_force	 \\
			subforlder "./external\_force" \\ 
		\end{tabular}		
		& Apply the gravity or body force to momentum equation \\	
		\hline	 
		\begin{tabular}{@{}c@{}} 
			general\_dynamics \\
			subforlder "./general\_dynamics" \\ 
		\end{tabular}		
		& 
		\begin{tabular}{@{}c@{}}
				Particle dynamics for applying periodic, in- and outflow BCs, \\
				implementing ghost-particle BCs and randomized particle position 
		\end{tabular} \\		
		\hline		
		\begin{tabular}{@{}c@{}} 
			fluid\_dynamics	 \\
			subforlder "./fluid\_dynamics" \\ 
		\end{tabular}		
		& 
		\begin{tabular}{@{}c@{}}
				Fluid particle dynamics for:\\
				1) Setting up the initial condition \\
				2) Computing advection and acoustic timestep size\\
				3) Computing particle number density \\
				4) Applying transport-velocity formulation \\
				5) Calculating viscous force \\
				6) Implementing the density initialization algorithm. \\
				7) Integrating momentum equation with position-based Verlet scheme
			\end{tabular}  \\		
		\hline
		\begin{tabular}{@{}c@{}} 
			solid\_dynamics		 \\
			subforlder "./solid\_dynamics	" \\ 
		\end{tabular}		
		& 
		\begin{tabular}{@{}c@{}} 
				Solid particle dynamics for :\\
				1) Setting up the initial condition \\
				2) Computing acoustic timestep size\\
				3) Computing or updating the normal direction of structure\\
				4) Computing the kernel correction matrix\\
				5) Computing pressure and viscous force form the fluid\\
				6) Integrating momentum equation with position-based Verlet scheme \\
				7) Computing total force on solid body for Simbody integration \\
				8) Constrain particle in solid body from Simbody integration state\\
			\end{tabular}  \\
		\hline
		\begin{tabular}{@{}c@{}} 
			diffusion\_reaction\_dynamics	\\
			subforlder "./diffusion\_reaction\_dynamics"
		\end{tabular}
	& 
		\begin{tabular}{@{}c@{}}
			Particle dynamics for : \\  
			1) Integrating the diffusion process with	RK2 scheme \\
			2) Integrating the reaction process with RBR splitting method 
		\end{tabular}  \\
	\hline
		\begin{tabular}{@{}c@{}} 
			active\_muscle\_dynamics \\
			subforlder "./active\_muscle\_dynamics" \\ 
		\end{tabular}		
	& 
		\begin{tabular}{@{}c@{}} 
				Computing electro-mechanical feedback
		\end{tabular}  \\					
	\hline	
	\end{tabular}
	\label{tab:particle-dynamics}
\end{table*}
%
%
\begin{table*}[htb!]
	\centering
	\caption{Program units (".h" and ".cpp") for the interface with Simbody in folder "SPHinxsys/src/share/simbody". }
	\begin{tabular}{ cc}
		\hline
		Program unit   &  Description \\ 
		\hline
		state\_engine & Interface for parsing state of MobilizedBody in Simbody  \\		
		xml\_engine  &  Interface for XML data parsing\\
		\hline	
	\end{tabular}
	\label{tab:simbody}
\end{table*}
\begin{table*}[htb!]
	\centering
	\caption{Program units (".h" and ".cpp") for the I/O system in folder "SPHinxsys/src/share/io\_system". }
	\begin{tabular}{ cc}
		\hline
		Program unit   &  Description \\ 
		\hline
		in\_output & \begin{tabular}{@{}c@{}} 
										Particle data read and write in XML format for restart \\ 
										Particle data output in PLT and VTU format for visualization \\
										Observed data output in DAT format for formal analysis
								\end{tabular} \\
		\hline	
	\end{tabular}
	\label{tab:io}
\end{table*}
%
\subsection{Installation and execution}\label{sec:code-compilation}
SPHinXsys source and executable files are distributed on a dedicated Git repository on GitHub. 
The executable files are released for cross-platform building, including Linux, MAC OSX and Microsoft Windows. 
SPHinxsys depends on the following libraries:
\begin{itemize}
	\item  cross-platform building: cmake 3.14.0 or later.
	\item compiler: Visual Studio 2017 or later (Windows only), gcc 4.9 or later (typically on Linux), or Apple Clang (1001.0.46.3)  or later
	\item BOOST library (newest version)
	\item TBB library (newest version)
	\item Simbody library 3.6.0 or later
	\item linear algebra: LAPACK 3.5.0 or later and BLAS 
\end{itemize}
The general procedure for installing and executing SPHinXsys is as follows:
\begin{itemize}
	\item installing the dependencies and set up the system variables in your machine as shown in Table \ref{tab:system-varialbe} 
	\item download the source code or clone the git repository
	\item create a directory in which the user will build a SPHinXsys project, e.g.  "~/simbody-build" 
	\item configure the SPHinXsys build with CMake
	\item compile and run the tests 
\end{itemize}
For more details, the readers are referred to the repository page \url{https://github.com/Xiangyu-Hu/SPHinXsys}
\begin{table*}[htb!]
	\centering
	\caption{System variables for dependencies search.}
	\begin{tabular}{ cc}
		\hline
		System variables   &  Path \\ 
		\hline
		TBB\_HOME & path/to/TBB-installation-prefix \\
		BOOST\_HOME & path/to/boost-installation-prefix \\
		SIMBODY\_HOME & path/to/Simbody-installation-prefix \\
		\hline	
	\end{tabular}
	\label{tab:system-varialbe}
\end{table*}
%
%
%
%
%
\section{Validations and applications}\label{sec:validation}
SPHinXsys has been validated and applied on more than $20$ test cases, 
whose main program are updated with the last code release. 
Some of them are briefly described in the following and also referred to the corresponding publications. 
The aim of this section is to recall the code validations and applications.
The references for details and validations on the single test cases are available in the following subsections, 
which are grouped according to the associated fields: 
fluid dynamics (Section \ref{sec:fluid-examples}), solid mechanics (Section \ref{sec:solid-examples}), 
FSI (Section \ref{sec:fsi-examples}), 
thermal and mass diffusion (Section \ref{sec:diffusion-examples}, 
reaction-diffusion (Section \ref{sec:ecr-examples}) and electromechanics (Section \ref{sec:emf-examples}).
\subsection{Fluid dynamics}\label{sec:fluid-examples}
In the context of modeling fluid dynamic, 
the WCSPH method is widely applied for the simulation of violent free-surface flows exhibiting violent events such as impact and breaking. 
Typical examples include dambreak flow, sloshing and wave impact, as well as fluid-solid interactions. 
Here, 
four benchmark tests involving violent free-surface flow are briefly summarized to validate SPHinXsys. 

\begin{figure*}
	\centering
	\begin{subfigure}[b]{0.4\textwidth}
		\includegraphics[trim = 10cm 5cm 10cm 5cm, clip,width=0.99\textwidth]{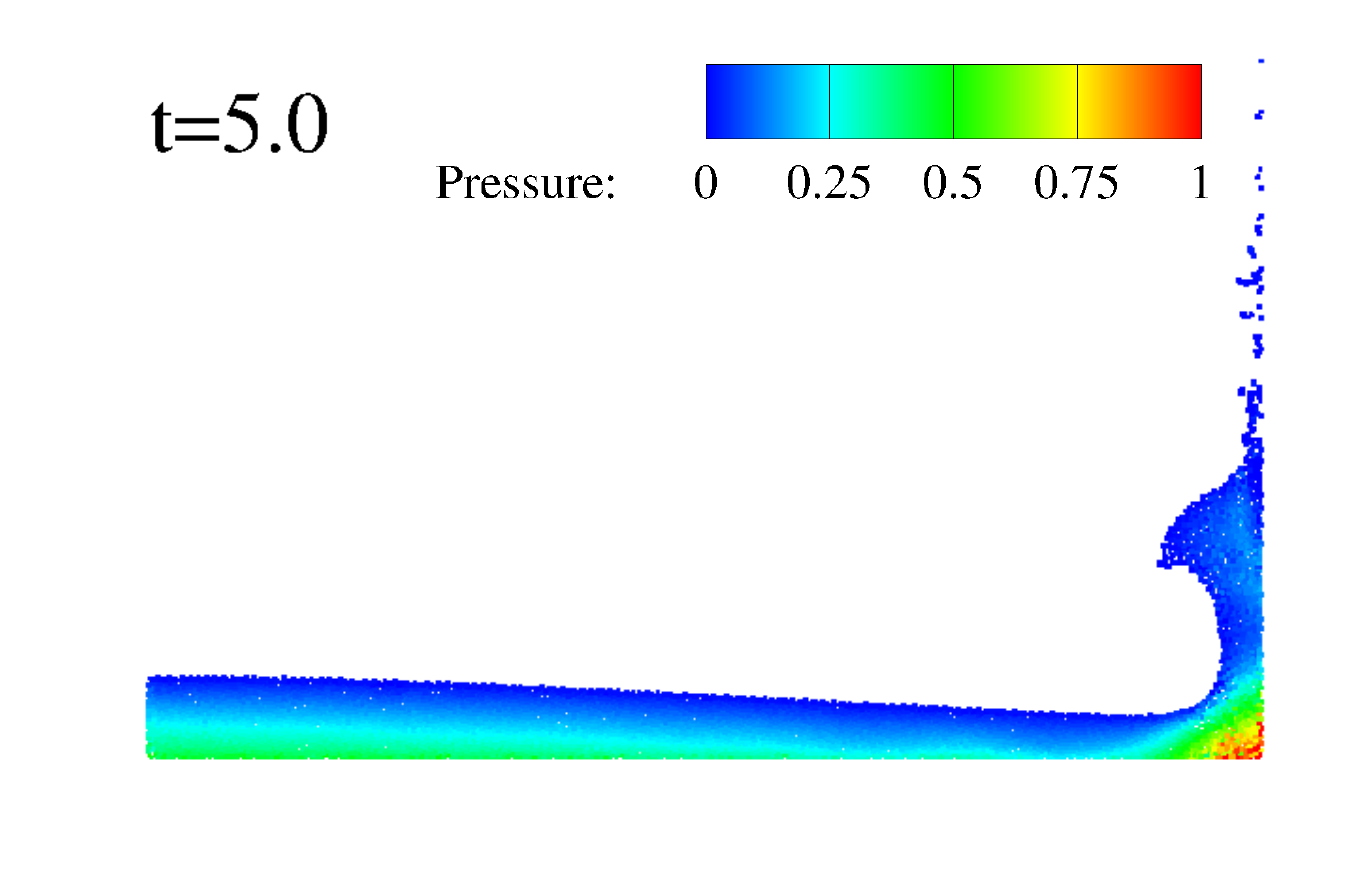} 
		\caption{2D dambreak flow}
		\label{fig:dam-2d}
	\end{subfigure}
	\begin{subfigure}[b]{0.59\textwidth}
		\vspace{0.5cm}
		\includegraphics[trim = 1mm 1mm 1mm 1mm, clip,width=0.99\textwidth]{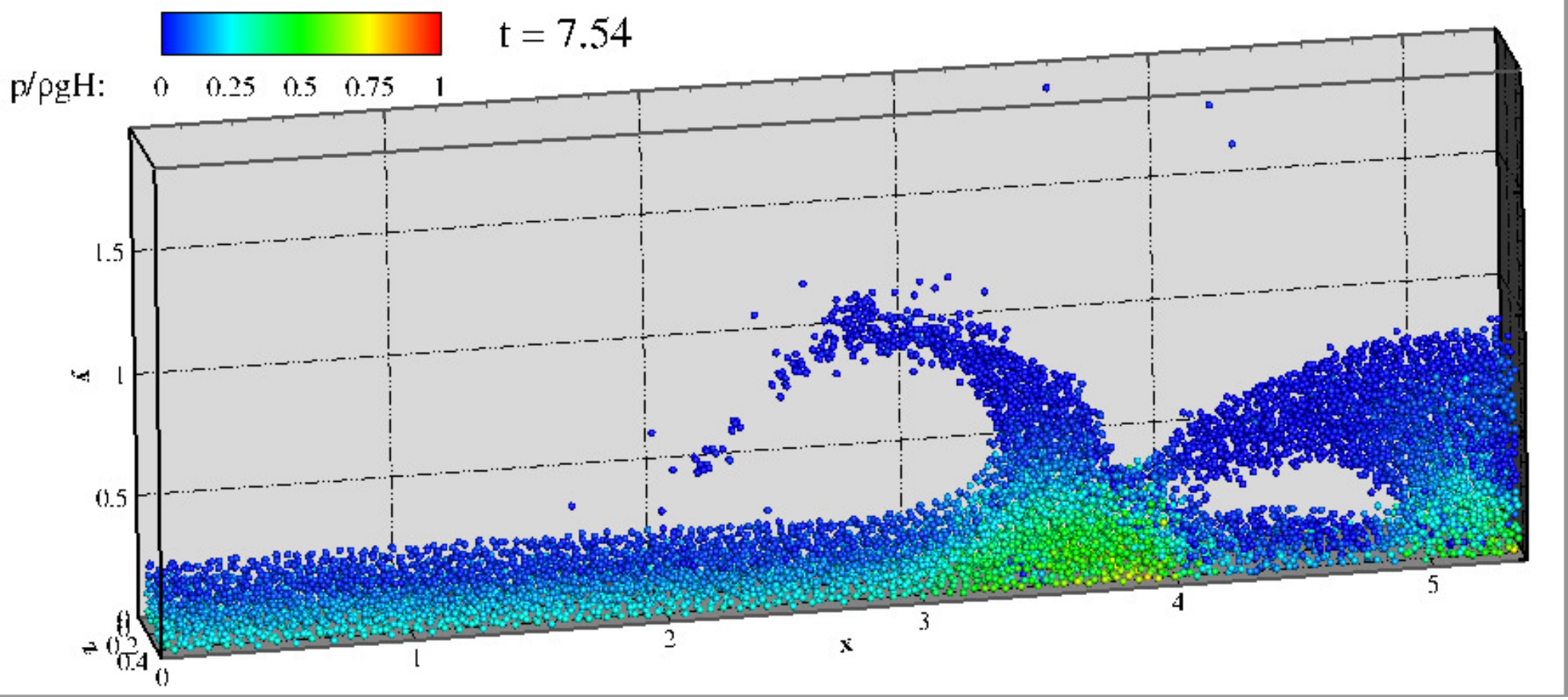}
		\caption{3D dambreak flow}
		\label{fig:dam-3d}
	\end{subfigure}
	\newline
	\begin{subfigure}[b]{0.59\textwidth}
		\vspace{0.5cm}
		\includegraphics[trim = 1mm 1mm 1mm 1mm, clip,width=0.99\textwidth]{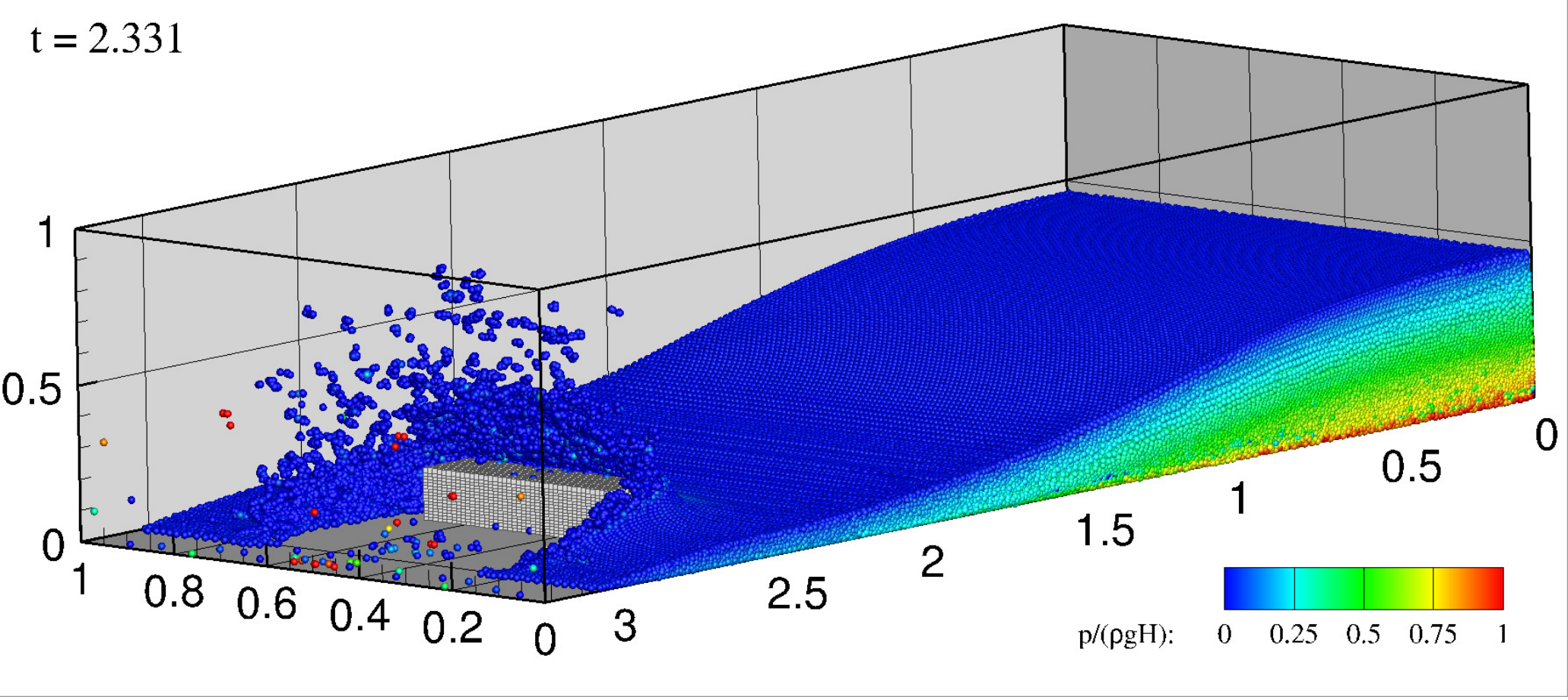}
		\caption{Dambreak flow with an obstacle}
		\label{fig:dam-obstacle}
	\end{subfigure}
	\begin{subfigure}[b]{0.4\textwidth}
		\vspace{0.5cm}
		\includegraphics[trim = 1mm 1mm 1mm 1mm, clip,width=0.99\textwidth]{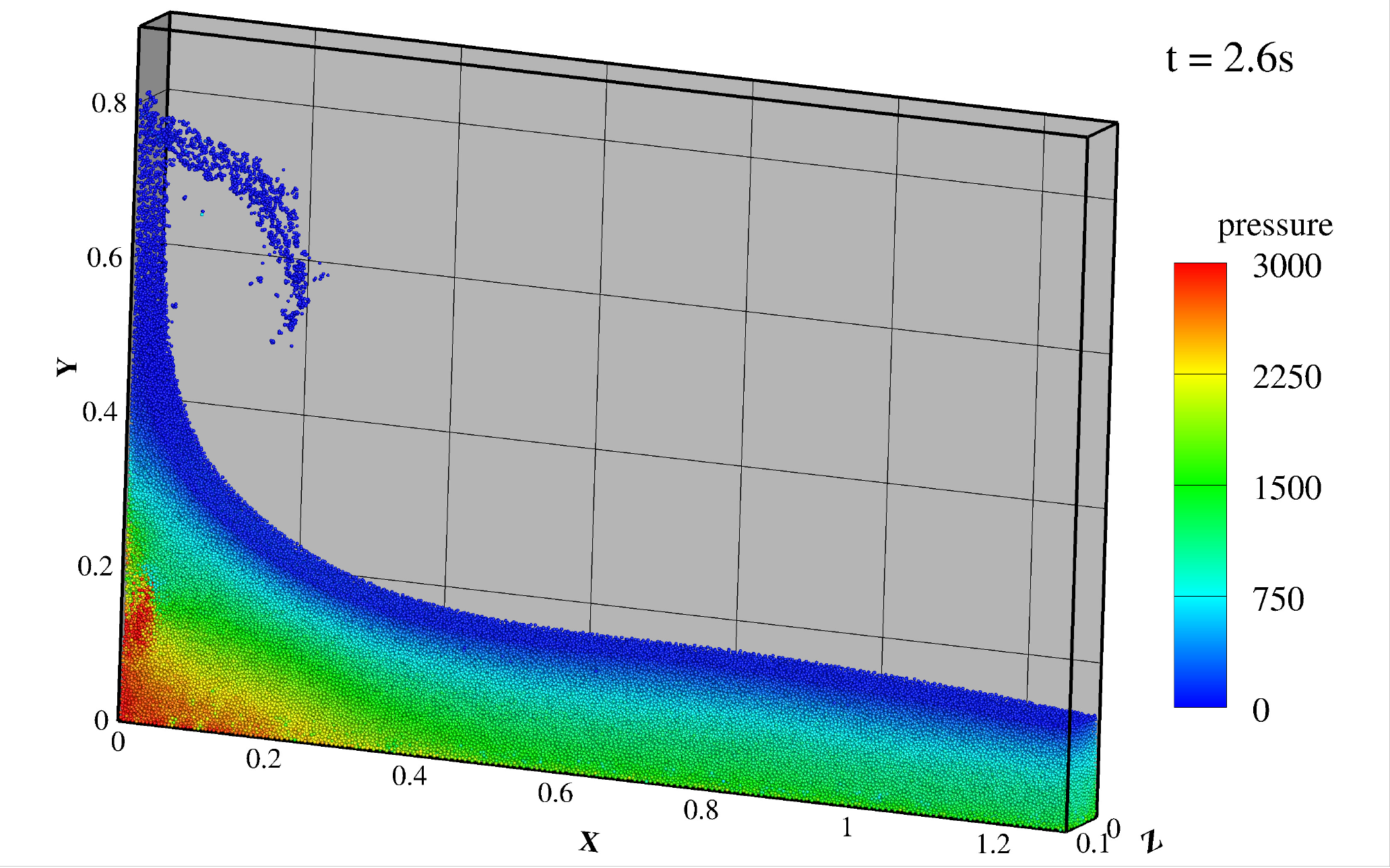}
		\caption{Sloshing tank}
		\label{fig:sloshing}
	\end{subfigure}
	\caption{SPHinXsys validations and applications for free-surface flows with violent wave-breaking and impact events.
		(a) Two-dimensional dambreak flow. 
		(b) Three-dimensional dambreak flow.
		(c) Three-dimensional dambreak flow impact on a obstacle.
		(d) Sloshing tank under resonance . }
	\label{fig:freesruface-flow}
\end{figure*}

The first two tests, 
viz. 2D and 3D dambreak flows 
(folders ‘‘cases\_test/test\_2d\_dambreak’’ and ‘‘cases\_test/test\_3d\_dambreak’’), 
allow us to quantitatively validate the program against the available experimental data. 
Figure \ref{fig:dam-2d} and \ref{fig:dam-3d} illustrate the snapshots of the free surface with the main features, 
including high roll-up along the downstream wall and a large reflected jet, being well captured. 
The third example simulates a dambreak flow interacting with a fixed obstacle and the solid boundaries of the domain as shown in Figure \ref{fig:dam-obstacle}. 
As another challenging problem, 3D sloshing tank 
has also been validated in SPHinXsys, which is introductory to the fuel/liquid natural gas (LNG) sloshing tank applications. 
Figure \ref{fig:sloshing} shows the particle and pressure distributions when a high run-up forms and impacts to the wall. 
For all the test cases, 
validation of the time history of impact pressure is documented by comparing the numerical results with the experimental data \cite{zhang2020dual}. 
%
\subsection{Solid dynamics}\label{sec:solid-examples}
SPHinXsys has been validated on preliminary benchmarks in 2D and 3D for solid dynamics where structure experiences large deformation. 

\begin{figure*}
	\centering
	\begin{subfigure}[b]{\textwidth}
		\centering
		\includegraphics[trim = 1mm 1mm 6cm  1mm, clip, width=0.495\textwidth]{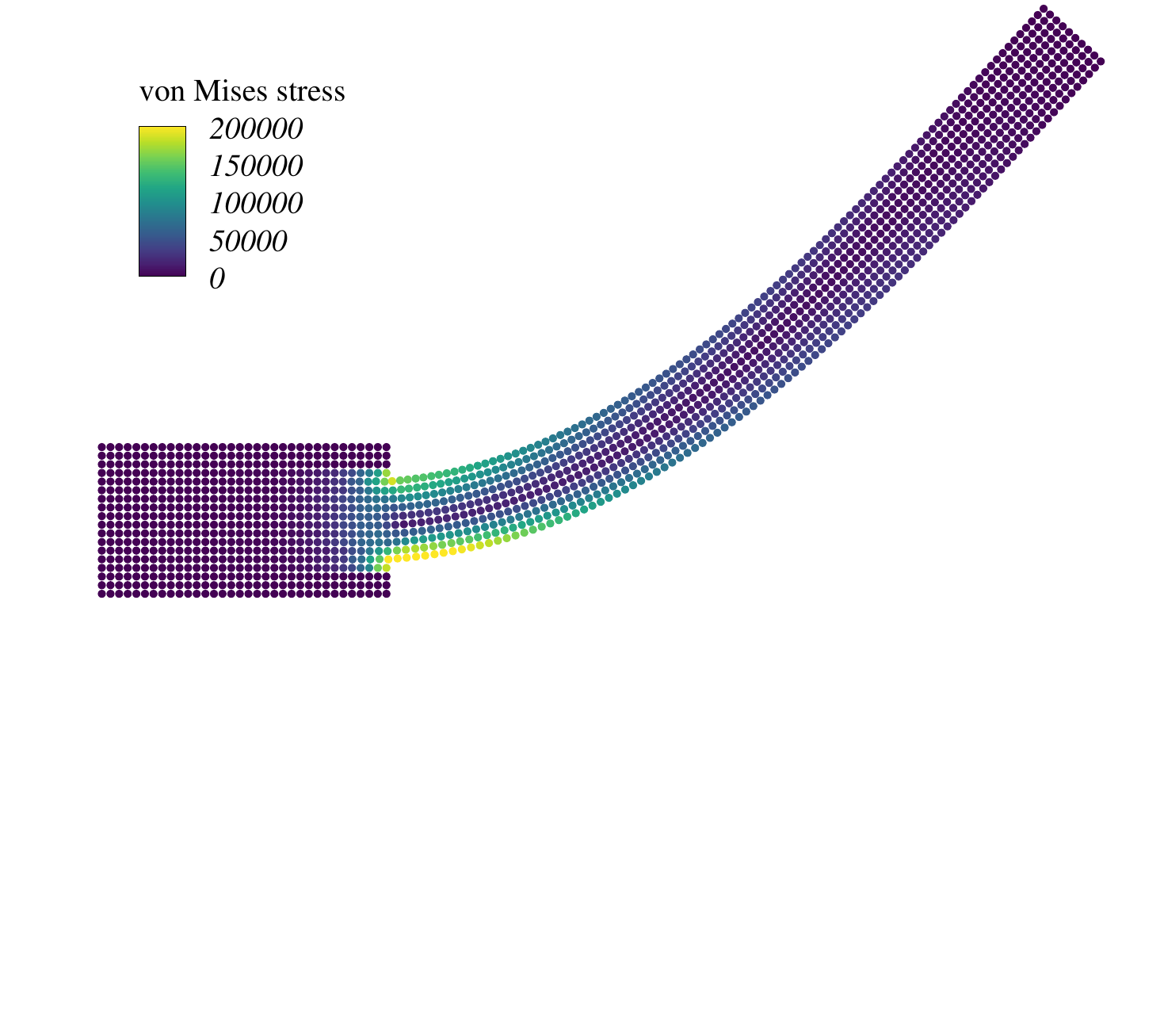}
		\includegraphics[trim = 1mm 1mm 6cm  1mm, clip, width=0.495\textwidth]{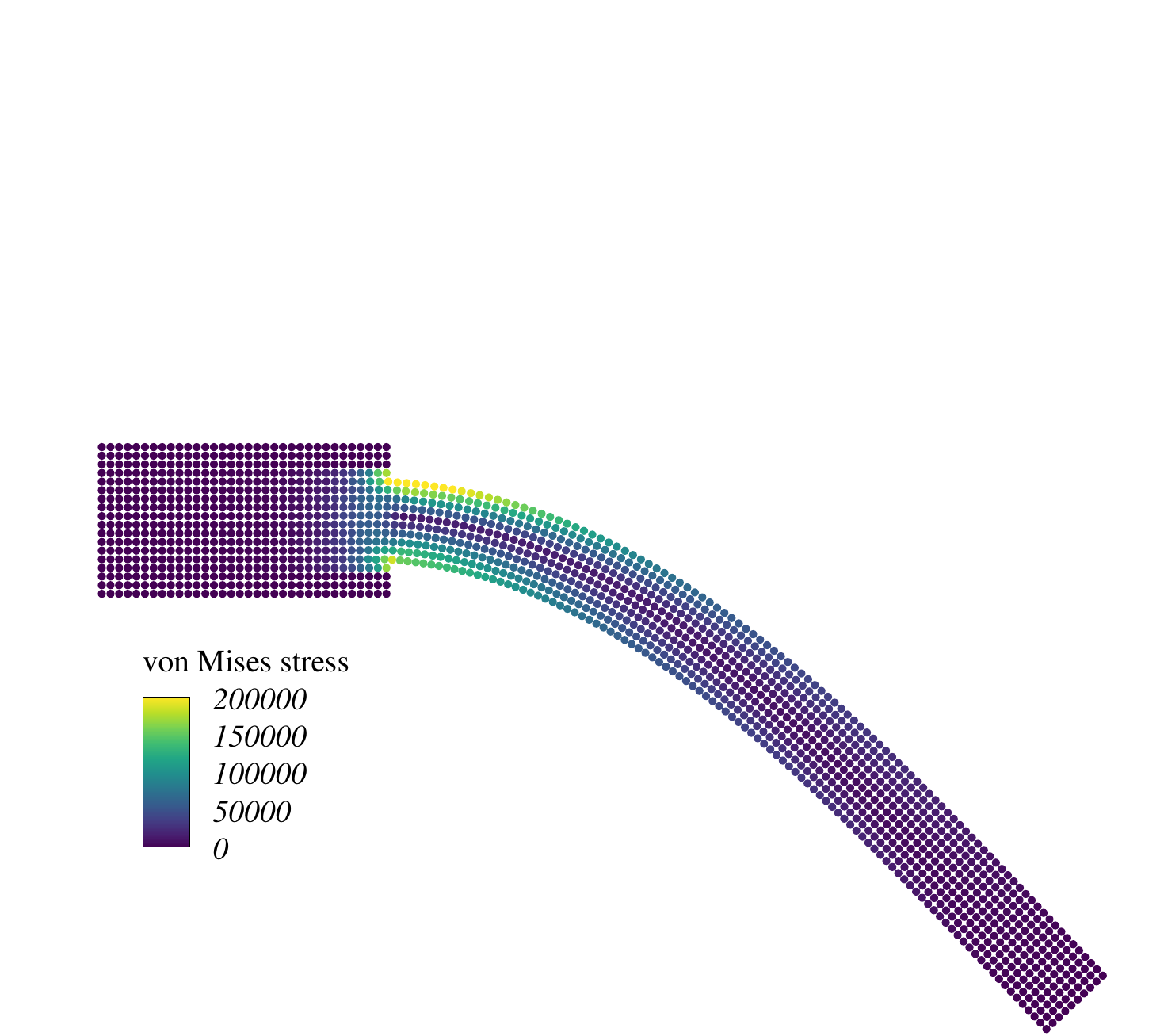}
		\caption{Oscillating beam}
		\label{fig:oscillating-beam}
	\end{subfigure}
	\newline 
	\begin{subfigure}[b]{\textwidth}
		\vspace{0.5cm}
		\includegraphics[trim = 1.75cm 1mm 9.5cm  1mm, clip, height=0.22\textwidth]{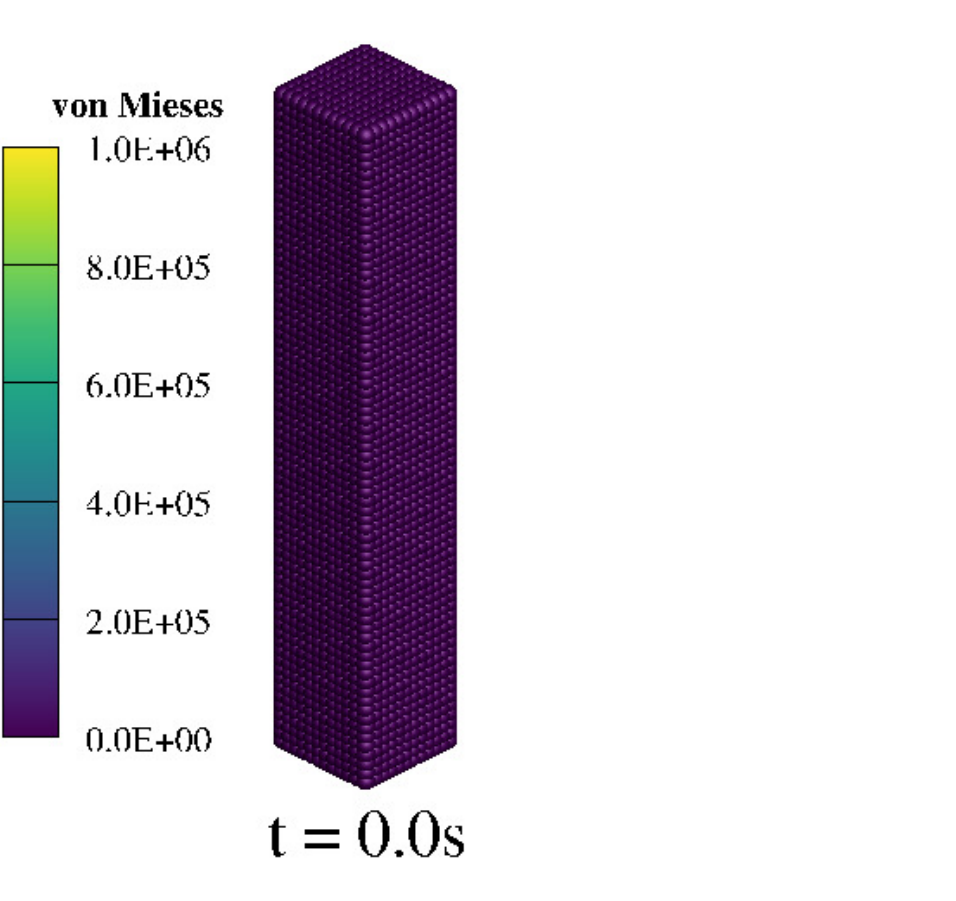}
		\includegraphics[trim = 4.5cm   1mm 9.5cm  1mm, clip, height=0.23\textwidth]{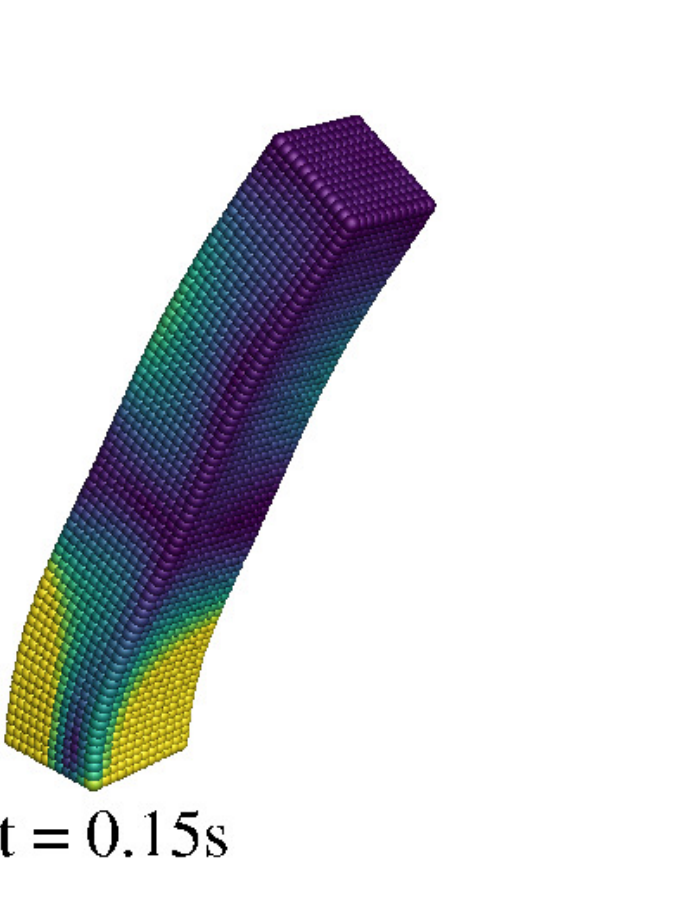}
		\includegraphics[trim = 4.5cm   1mm 9.5cm 1mm, clip, height=0.25\textwidth]{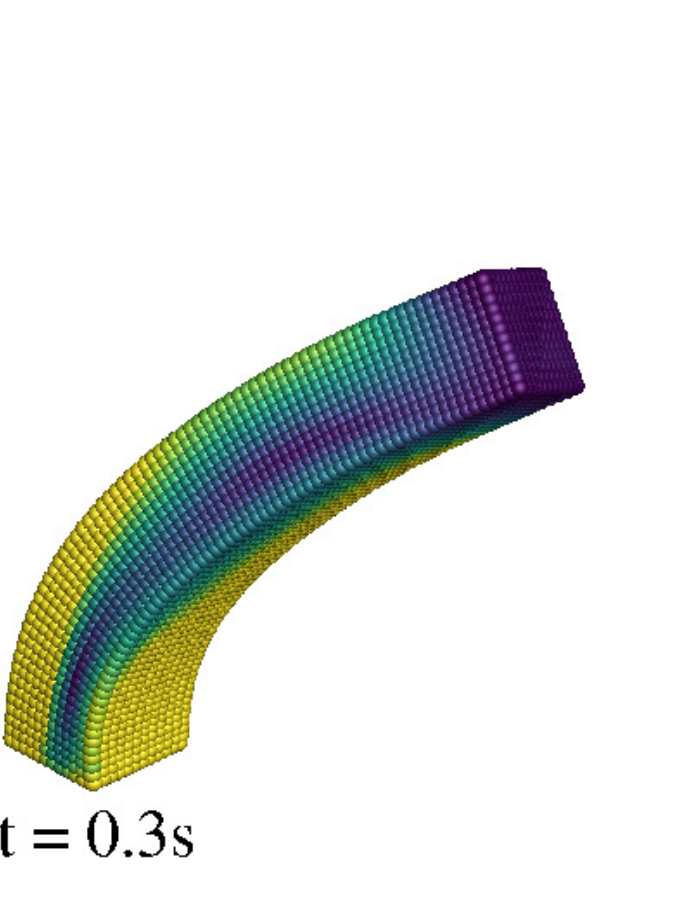}
		\includegraphics[trim = 4.5cm   1mm 9.5cm  1mm, clip, height=0.25\textwidth]{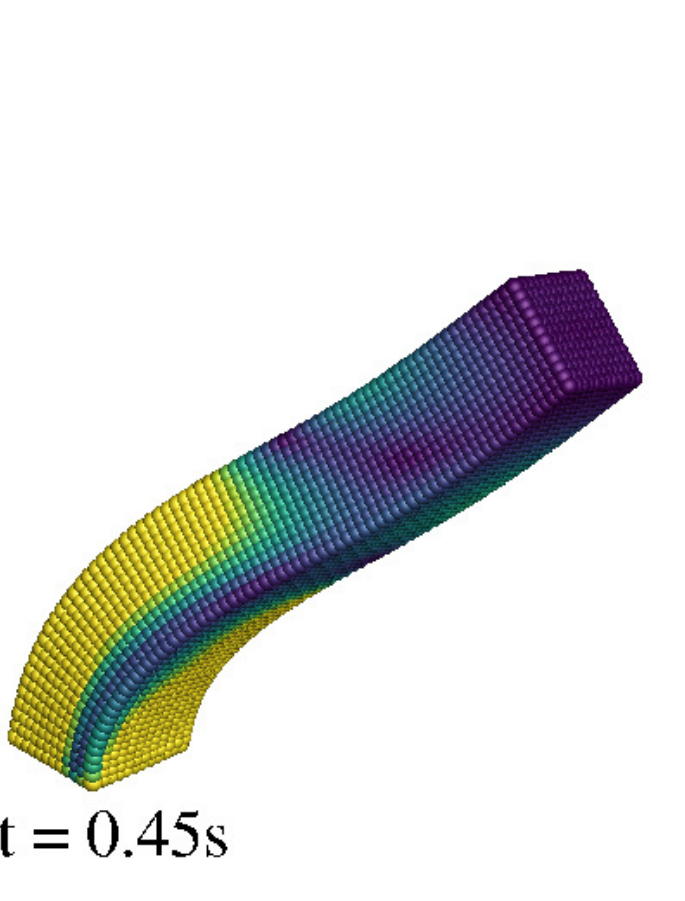}
		\includegraphics[trim = 4.5cm   1mm 9.5cm  1mm, clip, height=0.25\textwidth]{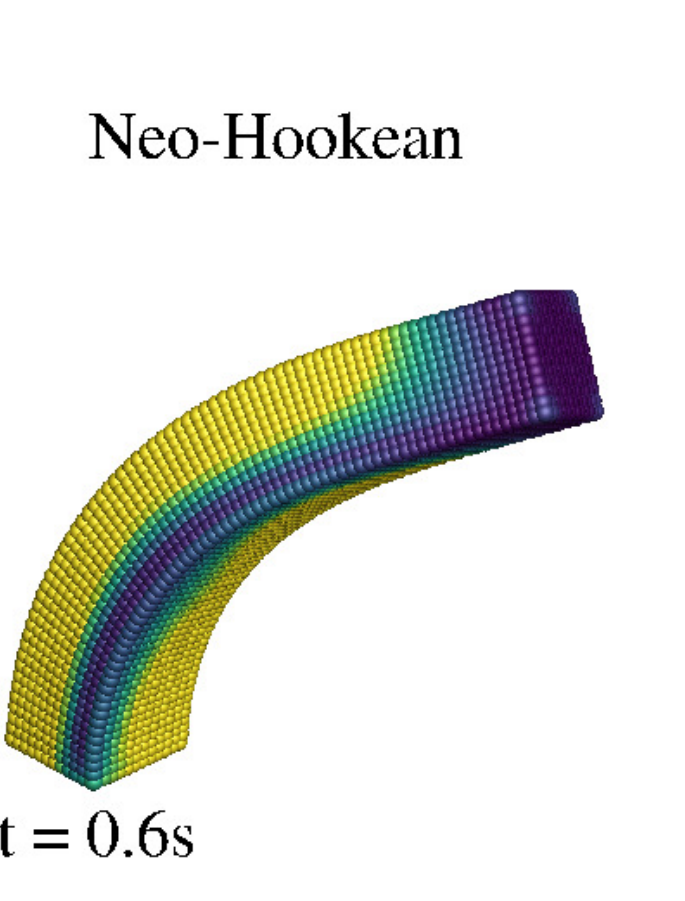} \\
		\caption{Cantilever}
		\label{fig:cantilever}
	\end{subfigure}
	\caption{SPHinXsys validations for the simulation of solid dynamics applications. 
		(a) Particle and von Mieses stress distributions in the deformed configuration with isotropic elastic model for 2D oscillating beam when the maximum deformation is reached. 
		(b) Particle and von Mieses stress distributions at different time instants in the deformed configuration with Holzapfel-Ogden model for a 3D bending cantilever.}
	\label{fig:soldidynamics}
\end{figure*}

The first benchmark, 
2D oscillating beam (folder "cases\_test/test\_2d\_oscillating\_beam") 
where a thin elastic beam initially stimulated by a velocity profile with one end fixed and another free, 
is investigated to demonstrate the numerical accuracy of solid mechanics solver in SPHinXsys. 
Figure \ref{fig:oscillating-beam} shows the particle and von Mises stress distribution when the beam reaches its maximum deformation. 
This test allows evaluating the accuracy for solid dynamics and code validation can be attained through the analytical solution.
In the second test, 
cantilever bending (folder "cases\_test/test\_3d\_cantilever") where a 3D rubber-like cantilever, 
whose bottom face is clamped to the ground and its
body is allowed to bend freely by imposing an initial uniform velocity, 
is considered. 
Figure \ref{fig:cantilever} shows the deformed configuration colored with von Mieses stress 
and for the quantitative comparison with data in literature the reader is referred to Ref. \cite{zhang2020integrative}.
%
\subsection{Fluid-structure interactions}\label{sec:fsi-examples}
In this part, 
three benchmark FSI tests, 
viz. 
a hydrostatic water column on an elastic plate, 
a flow-induced vibration and a dambreak flow with elastic gate, 
are studied in multi-resolution scenarios to validate the FSI solvers in SPHinXsys. 

\begin{figure*}
	\centering
	\begin{subfigure}[b]{0.45\textwidth}
		\includegraphics[trim = 2mm 2mm 2mm 2mm, clip,width=0.99\textwidth]{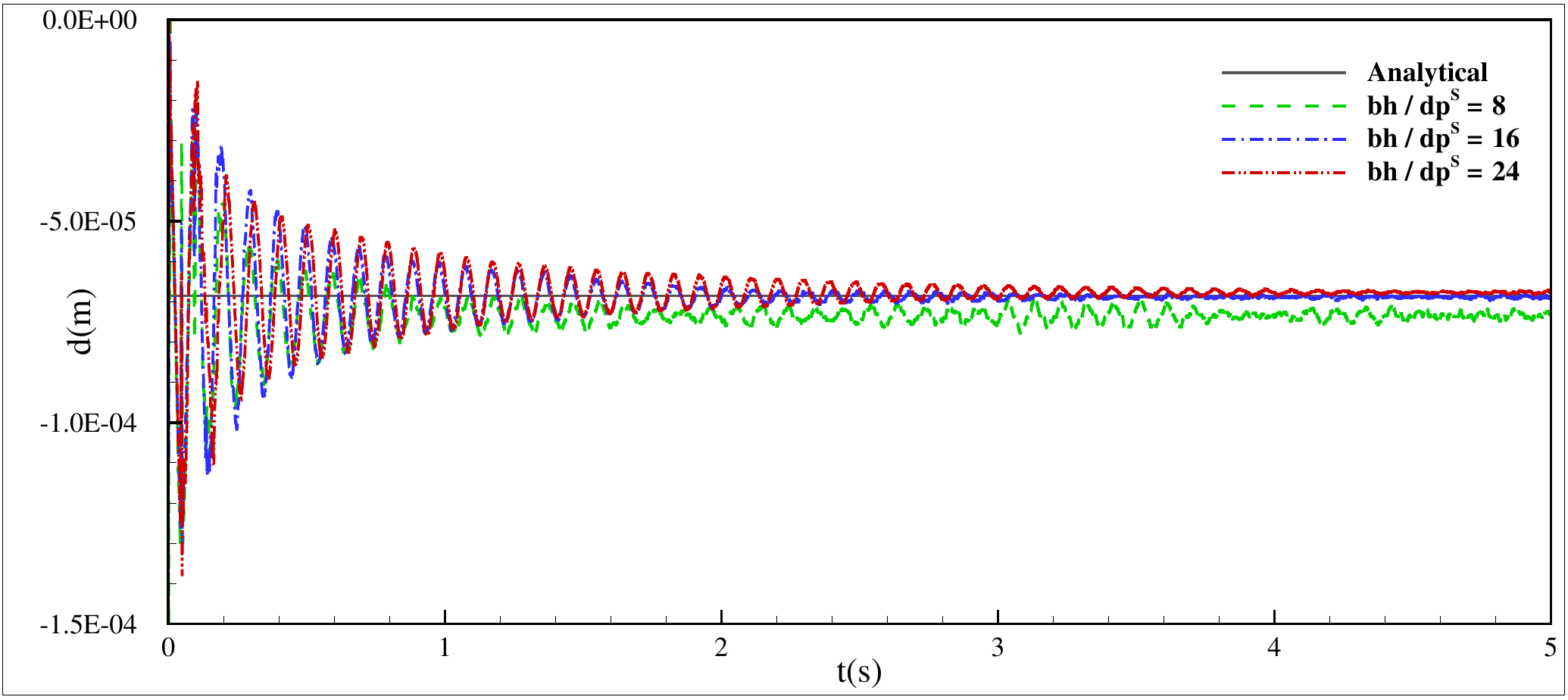} 
		\caption{Plate under hydrostatic water}
		\label{fig:hydrostatic}
	\end{subfigure}
	\begin{subfigure}[b]{0.54\textwidth}
		\vspace{0.5cm}
		\includegraphics[trim = 5mm 1cm 2mm 1cm, clip,width=0.99\textwidth]{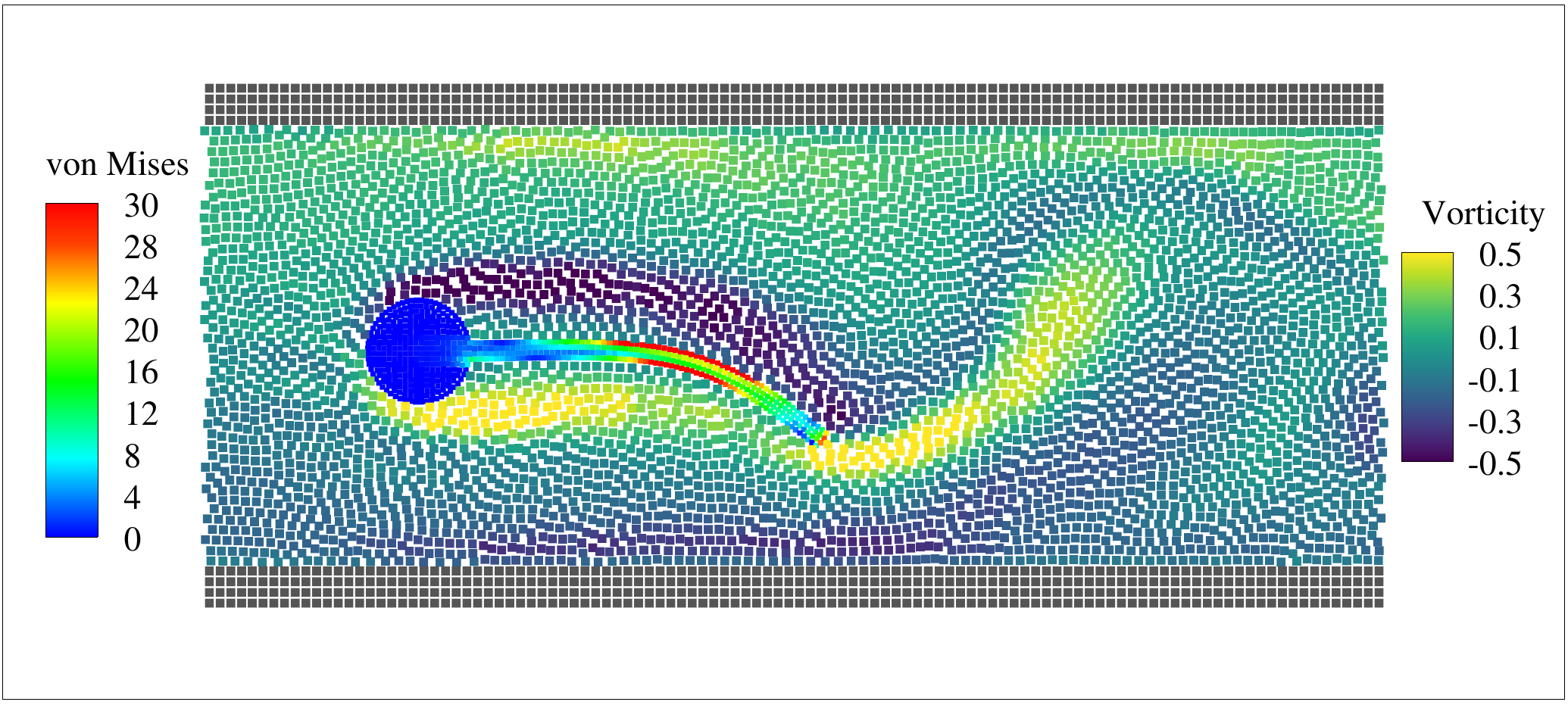}
		\caption{Flow-induced vibration}
		\label{fig:fsi}
	\end{subfigure}
	\newline
	\begin{subfigure}[b]{\textwidth}
		\vspace{0.5cm}
		\includegraphics[trim = 5mm 2.5mm 2mm 1cm, clip,width=\textwidth]{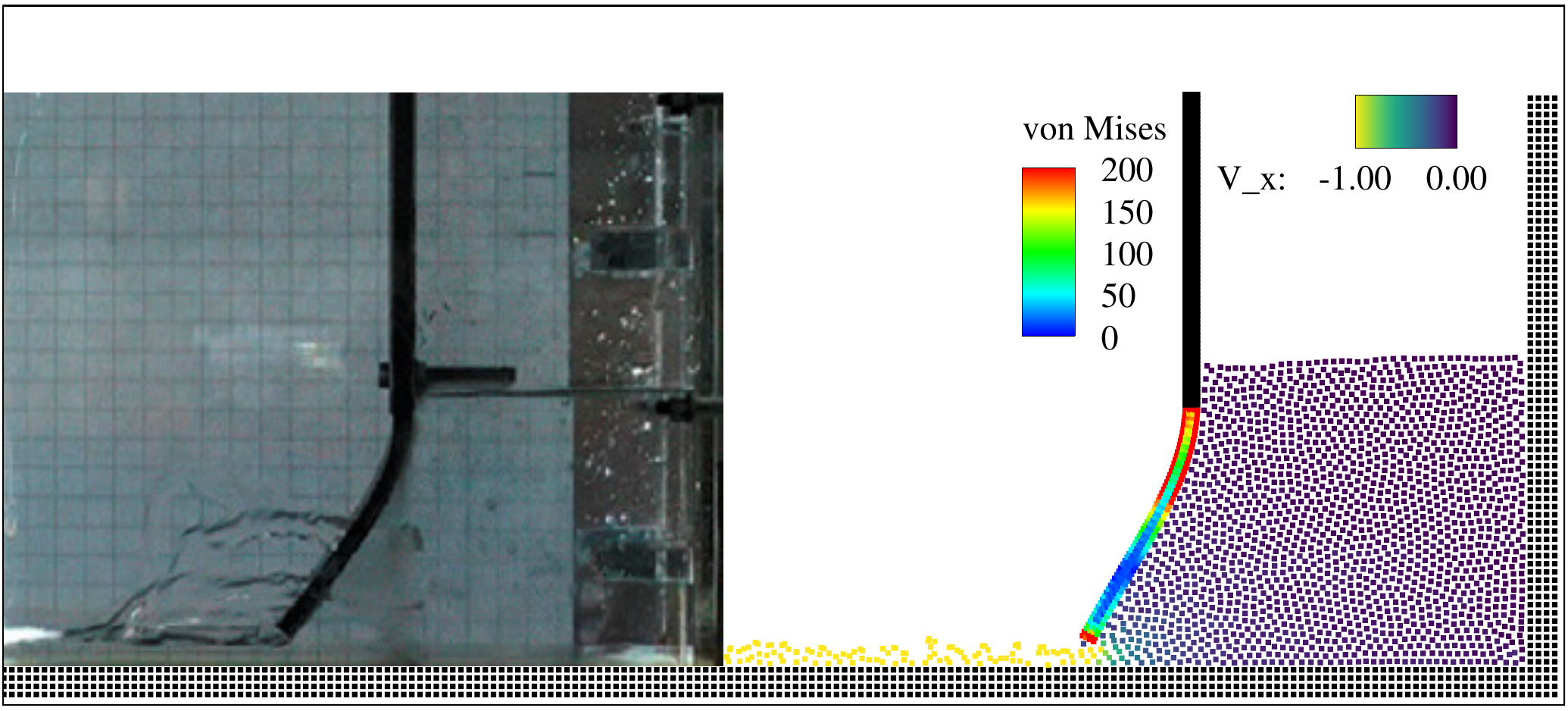}
		\caption{Dam-break flow through an elastic gate}
		\label{fig:dam}
	\end{subfigure}
	\caption{SPHinXsys validations and applications for FSI with a fluid-structure resolution ratio of $2$. 
		(a) The time history of mid-span displacement of an elastic plate under a hydrostatic water column. 
		(b) The flow vorticity field and beam deformation of flow-induced vibration of a beam attached to a cylinder.
		(c) Snapshots for dam-break flow through an elastic gate compared against experimental frames \cite{antoci2007numerical}. }
	\label{fig:mr-sph}
\end{figure*}

In the first test,  plate deformation under hydrostatic pressure of a water column is considered.
Figure \ref{fig:hydrostatic} gives the time histories of the mid-span displacement, together with the convergence 
study with increasing spatial resolution of structure while the resolutions of fluid is constant. 
A high-order convergence of the middle-span displacement is observed. 
In the second benchmark (folder "cases\_test/test\_2d\_fsi") , a two dimensional flow-induced vibration of a flexible beam attached to a rigid cylinder is studied. 
Figure \ref{fig:fsi} shows the flow vorticity field and beam deformation when self-sustained oscillation is reached. 
Furthermore, we consider the deformation of an elastic plate subjected to a
time-dependent water pressure (folder  "cases\_test/test\_2d\_dambreak\_gate").
The comparison between the numerical snapshots and the experiment presented by Antoci et al. \cite{antoci2007numerical} is illustrated in Figure \ref{fig:dam}. 
It can be observed that the simulation results are in a good agreement with the experimental results. 

Concerning the computational efficiency, with the multi-resolution treatments of SPHinXsys
up to 240 and 960 times speed ups are achieved when the fluid-structure resolution ratio is 2 and 4, respectively.
These computational efficiency analysis are carried out for the first benchmark. 
%
\subsection{Mass diffusion}\label{sec:diffusion-examples}
In this section, 
benchmark tests with available analytical solutions are investigated for validating the thermal and mass diffusion solvers in SPHinXsys. 

\begin{figure*}
	\centering
	\begin{subfigure}[b]{0.95\textwidth}
		\includegraphics[trim = 1mm 1mm 1mm 1mm, clip,width=0.45\textwidth]{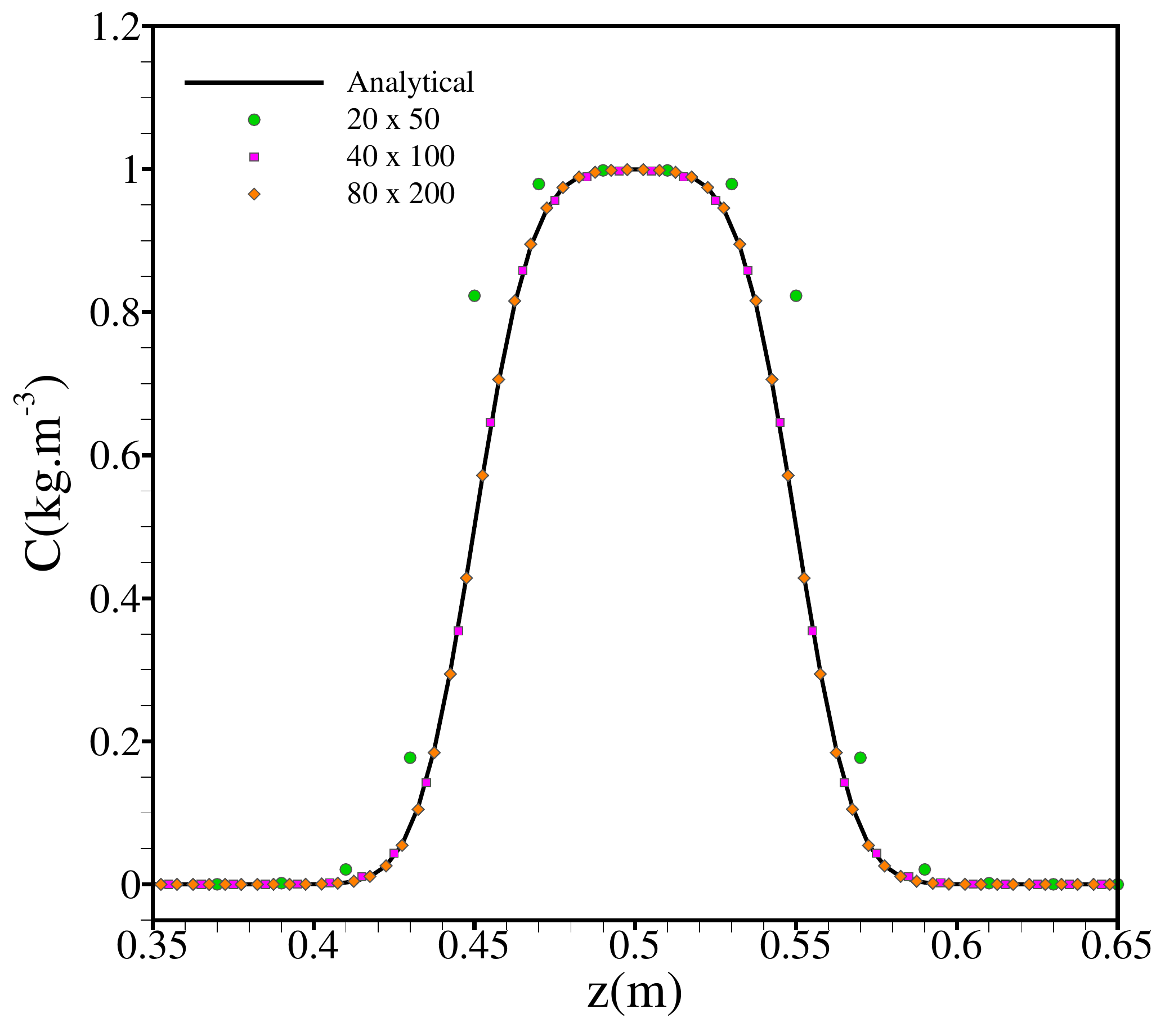}
		\includegraphics[trim = 1mm 1mm 1mm 1mm, clip,width=0.45\textwidth]{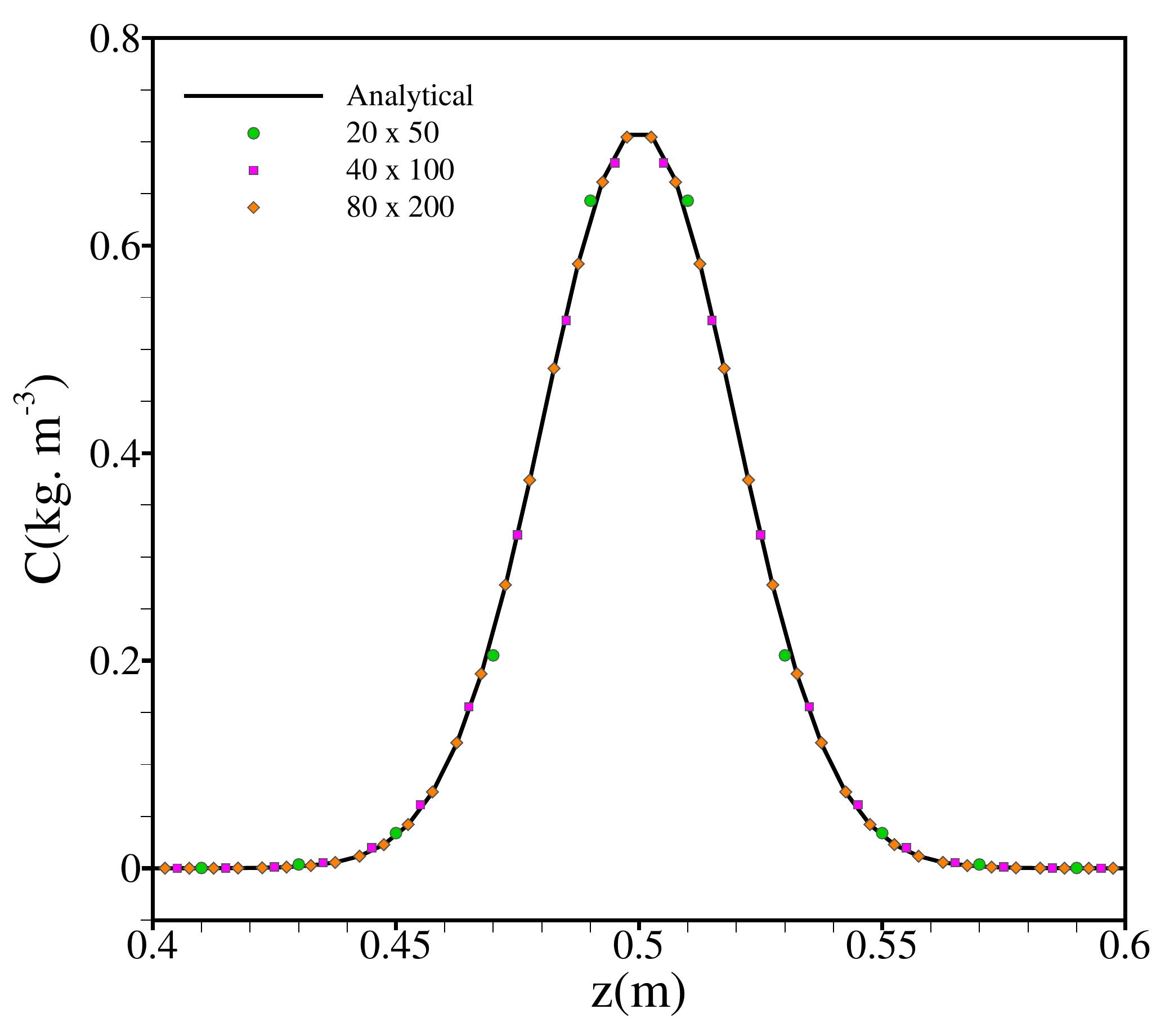}
		\caption{Isotropic diffusion with constant (left panel) and exponential (right panel) initial distributions.}
		\label{fig:iso-diffusion}
	\end{subfigure}
	\newline
	\begin{subfigure}[b]{\textwidth}
		\vspace{0.5cm}
		\includegraphics[trim = 1mm 3cm 1mm 1cm, clip,width=\textwidth]{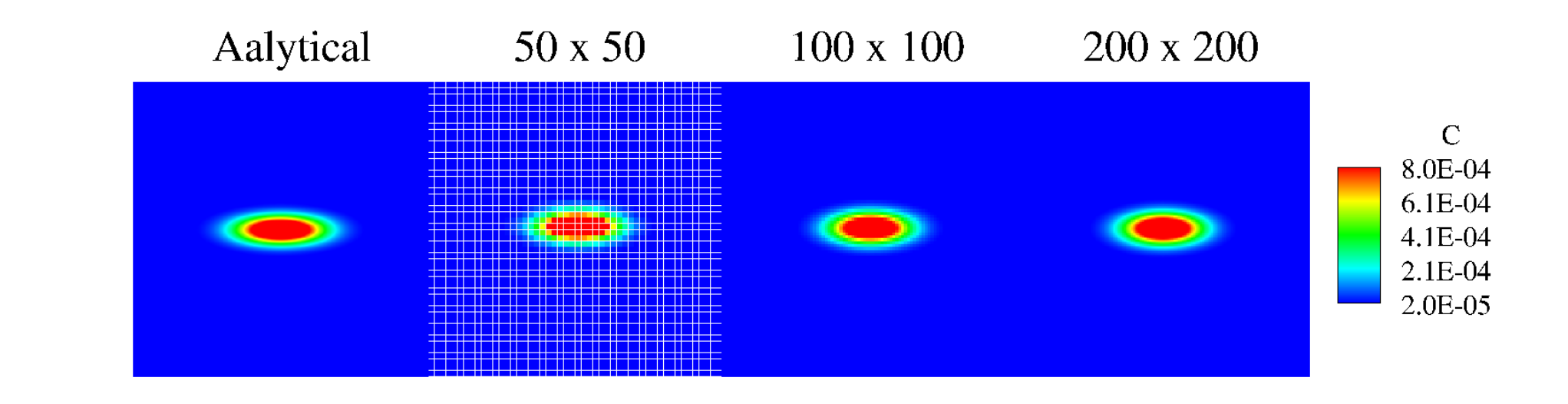}
		\caption{Concentration distributions for anisotropic diffusion.}
		\label{fig:aniso-diffusion}
	\end{subfigure}
	\newline
		\begin{subfigure}[b]{0.95\textwidth}
		\includegraphics[trim = 1mm 1mm 1mm 1mm, clip,width=0.45\textwidth]{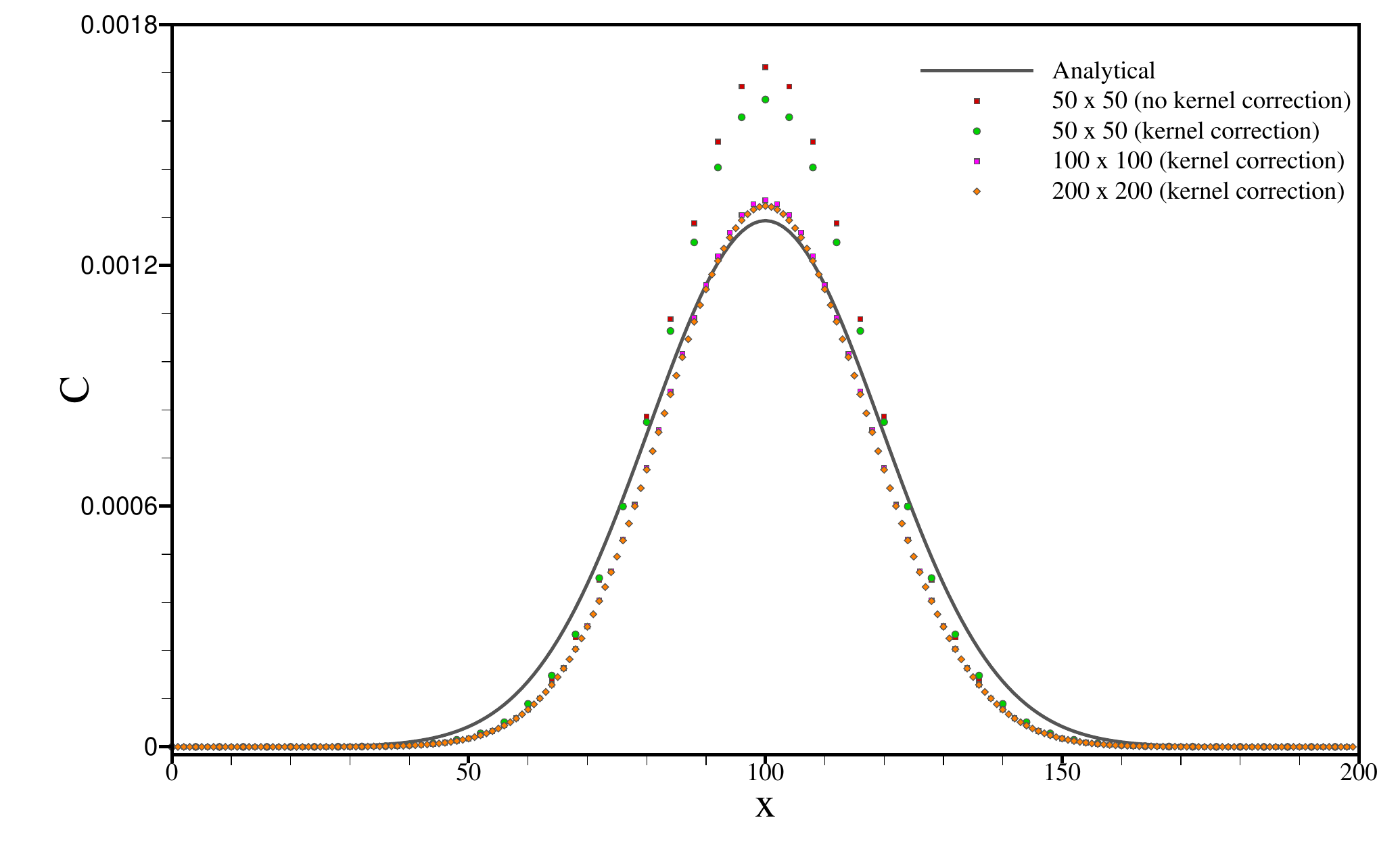} 
		\includegraphics[trim = 1mm 1mm 1mm 1mm, clip,width=0.45\textwidth]{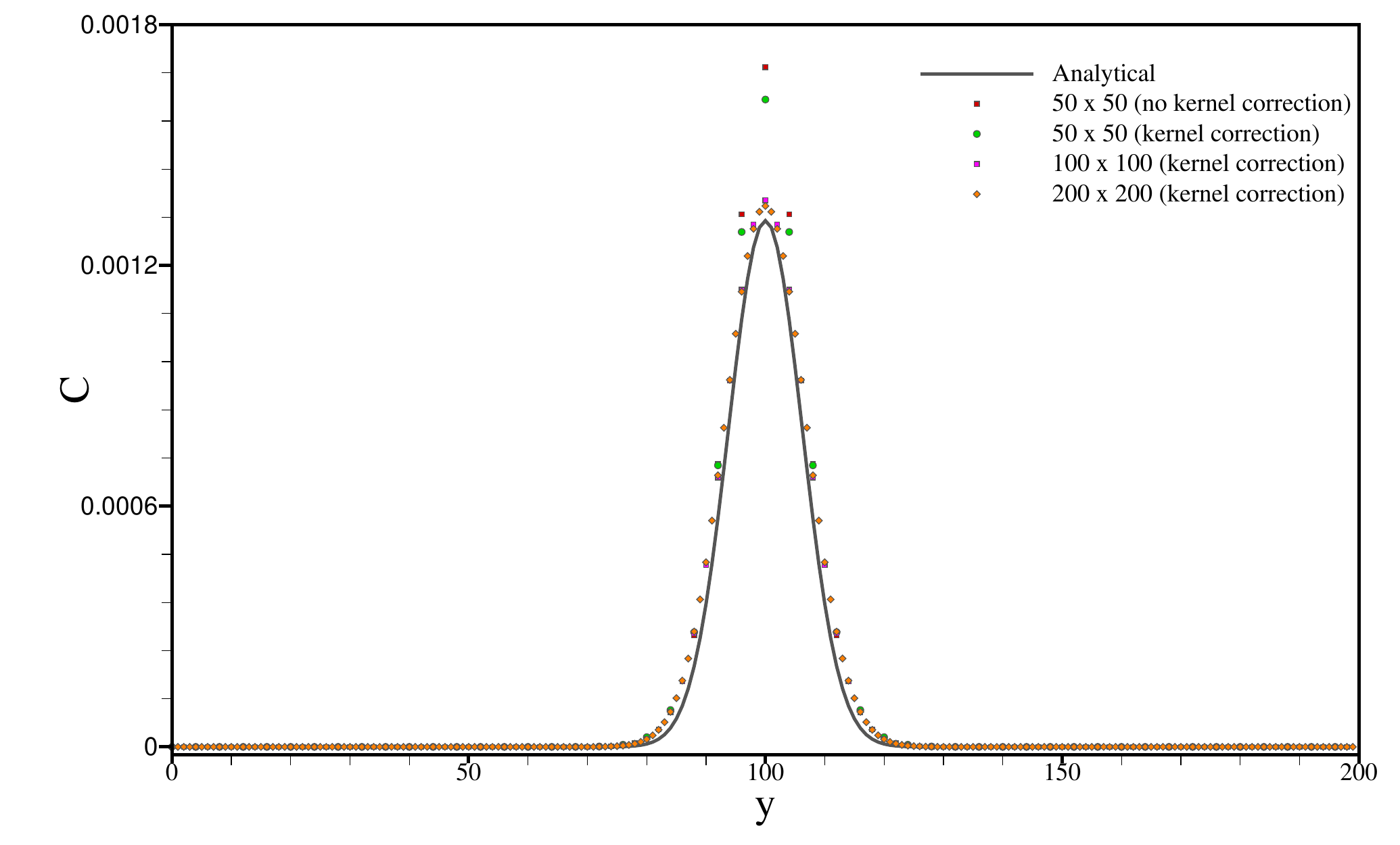}
		\caption{Concentration distributions in $x$, and $y$-direction for anisotropic diffusion.}
		\label{fig:aniso-diffusion-data}
	\end{subfigure}
	\caption{SPHinXsys validations for thermal diffusion with iso- and aniso-tropic diffusion tensor. 
		(a) Concentration distribution for isotropic diffusion processes with constant and exponential initial concentration distributions.
		(b) Concentration distribution for anisotropic diffusion with anisotropic ratio equals $10$.
		(c) Concentration distributions at horizontal cross in $x$, and $y$-direction for anisotropic diffusion. }
	\label{fig:diffusion}
\end{figure*}

The first test (folder "cases\_test/test\_2d\_diffusion") studied herein considers a 1D isotropic diffusion rectangle filled with water and a finite horizontal band of pollutant located in the middle of the rectangle. 
The initial conditions with both a constant and an exponential pollutant concentration distribution are considered. 
Figure \ref{fig:iso-diffusion} illustrates the comparison of the present predictions of the concentration distributions against the analytical solution. 
The second example considers an anisotropic diffusion process from a contaminant source in water, 
where the contaminant source is located in a two dimensional square computation domain and a higher anisotropic ratio is considered. 
Figure \ref{fig:aniso-diffusion} shows the computational concentration distribution and the comparison with analytical solution. 
Figure \ref{fig:aniso-diffusion-data} gives the numerical concentration distributions at horizontal cross section and vertical cross section and the corresponding analytical solution. 
It can be noted that SPHinXsys can accurately predict the concentration distribution in iso- and ansiotropic diffusion processes. 
%
\subsection{Diffusion-reaction equation}\label{sec:ecr-examples}
In this part, 
SPHinXsys is validated for solving reaction-diffusion model by capturing the free-pulse propagation of transmembrane potential 
and reproducing the spiral wave in uniform and nonuniform computational domains. 

\begin{figure*}
	\centering
	\begin{subfigure}[b]{\textwidth}
		\centering
		\includegraphics[trim = 2mm 2mm 2mm 2mm, clip, width=0.45\textwidth]{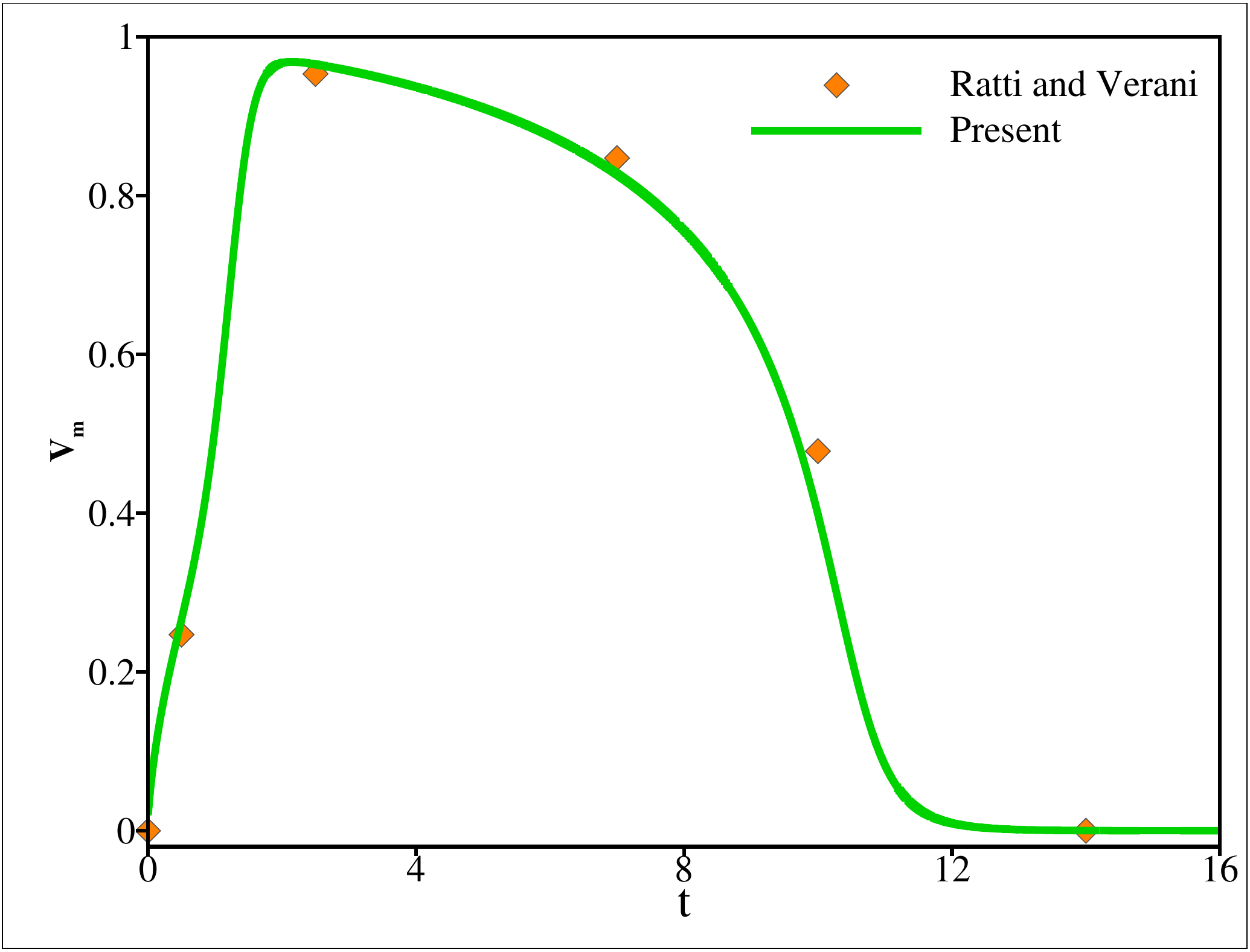}
		\caption{Depolarization of transmembrane potential.}
		\label{fig:depolarization}
	\end{subfigure}
	\newline
	\begin{subfigure}[b]{0.485\textwidth}
		\vspace{0.5cm}
		\includegraphics[trim = 5mm 25mm 5mm 15mm, clip, width=\textwidth]{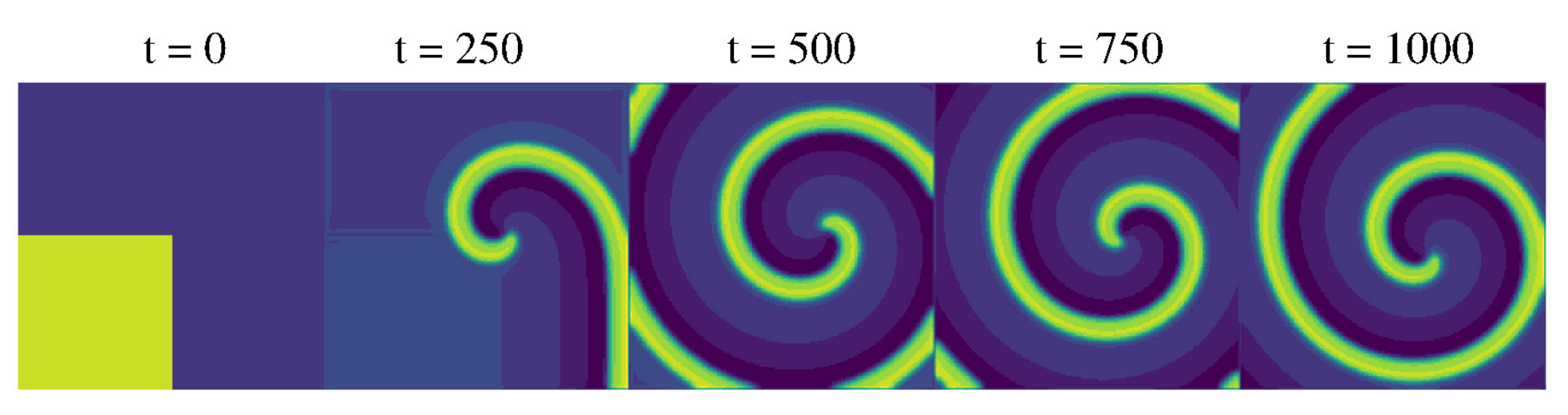}\\
		\includegraphics[trim = 5mm 25mm 5mm 25mm, clip, width=\textwidth]{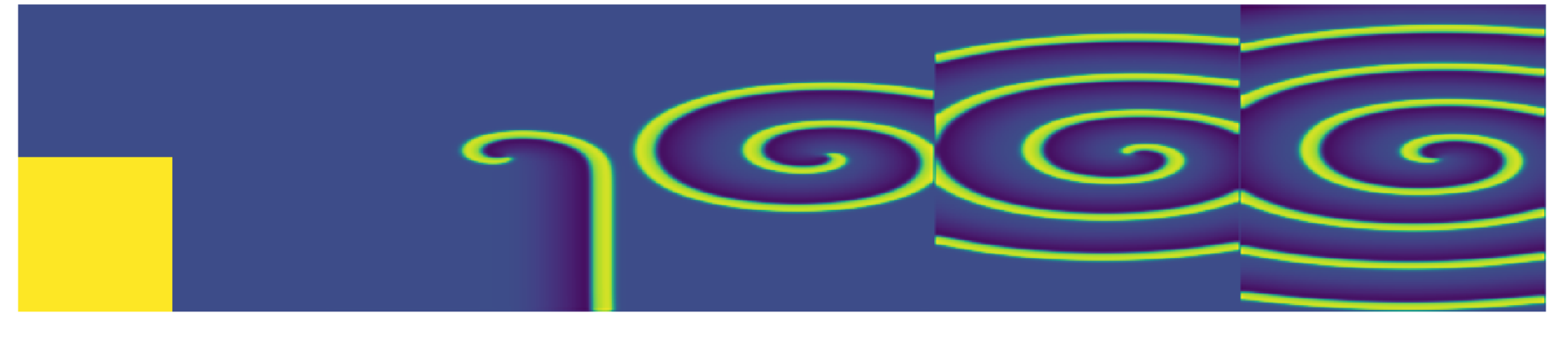}
		\caption{Spiral wave in rectangular geometry.}
		\label{fig:spiralwave-square}
	\end{subfigure}
	\begin{subfigure}[b]{0.485\textwidth}
		\vspace{0.5cm}
		 \includegraphics[trim = 5mm 25mm 5mm 15mm, clip, width=\textwidth]{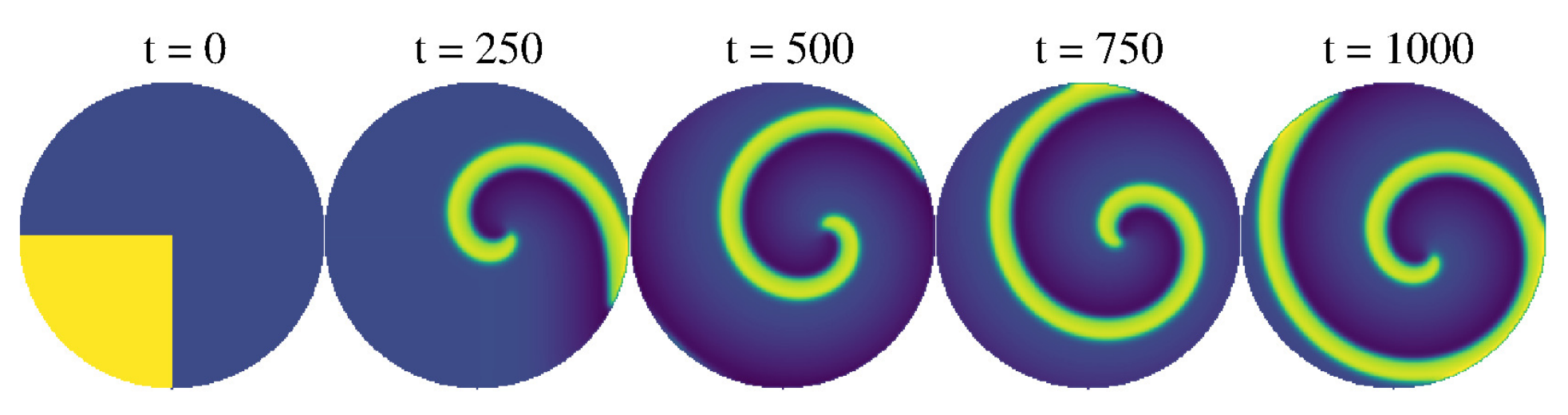}\\
		\includegraphics[trim = 5mm 25mm 5mm 25mm, clip, width=\textwidth]{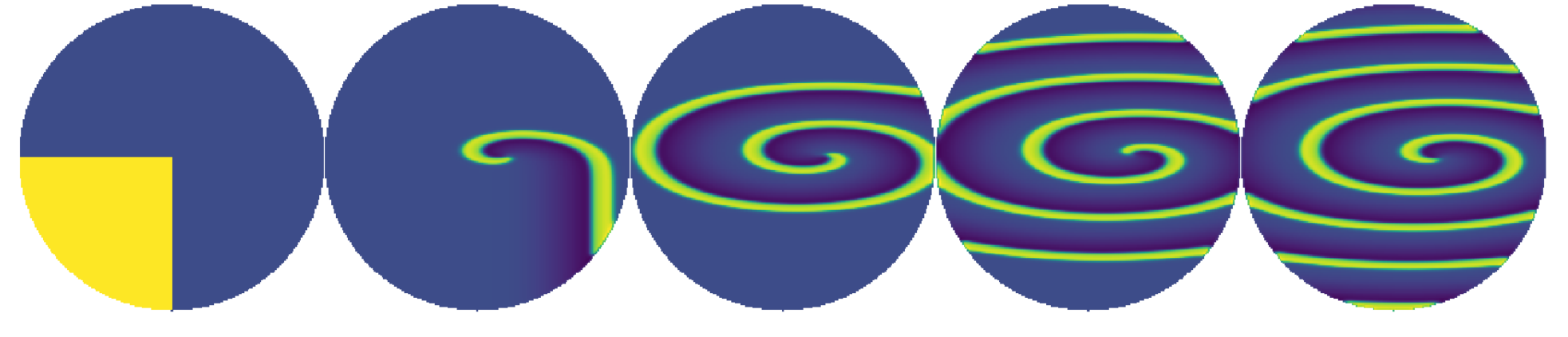}
		\caption{Spiral wave in circular geometry.}
		\label{fig:spiralwave-circle}
	\end{subfigure}
	\caption{SPHinXsys validations and applications for reaction-diffusion equation (monodomain equation). 
		(a) Time evolution of the transmembrane potential at point (0.3,0.7) in comparison with the results reported by Ratti and Verani \cite{ratti2019}. 
		(b) Spiral waves of the FitzHugh-Nagumo model in rectangular computational domain with iso- and anisotropic diffusion tensor.
		(c) Spiral waves of the FitzHugh-Nagumo model in circular computational domain with iso- and anisotropic diffusion tensor. }
	\label{fig:diffusion-reaction}
\end{figure*}

The first benchmark test (folder "cases\_test/test\_2d\_depolarization") 
taken from Ratti and Verani \cite{ratti2019}  
considers a transmembrane potential which propagates in a 2D isotropic tissue in a square domain. 
Figure \ref{fig:depolarization} reports the predicted evolution profile of the transmembrane potential
and the corresponding comparison with that of Ratti and Verani \cite{ratti2019}. 
It is noted that in accordance with the previous numerical estimation and experimental observation, 
the quick propagation of the stimulus in the tissue and the slow decrease in the transmembrane potential after a plateau phase are well predicted.
The second example considers the generation of an spiral wave in rectangular and circular geometries.  
Figure \ref{fig:spiralwave-square} and \ref{fig:spiralwave-circle} 
show spiral waves of the stable rotation solution in rectangular and circular computational domain at different time instants. 
Also, both iso- and anisotropic diffusion coefficient tensors are taken into consideration. 
As expected, the spiral wave generates a curve and rotates clockwise as reported
%
\subsection{Electromechanics}\label{sec:emf-examples}
In this section, 
the validation of SPHinXsys in modeling electromechanics is presented and the application for biventricular heart model is also summarized.  

\begin{figure*}
	\centering
	\begin{subfigure}[b]{\textwidth}
		\centering
		\includegraphics[trim = 2mm 2mm 2mm 2mm, clip, width=0.6\textwidth]{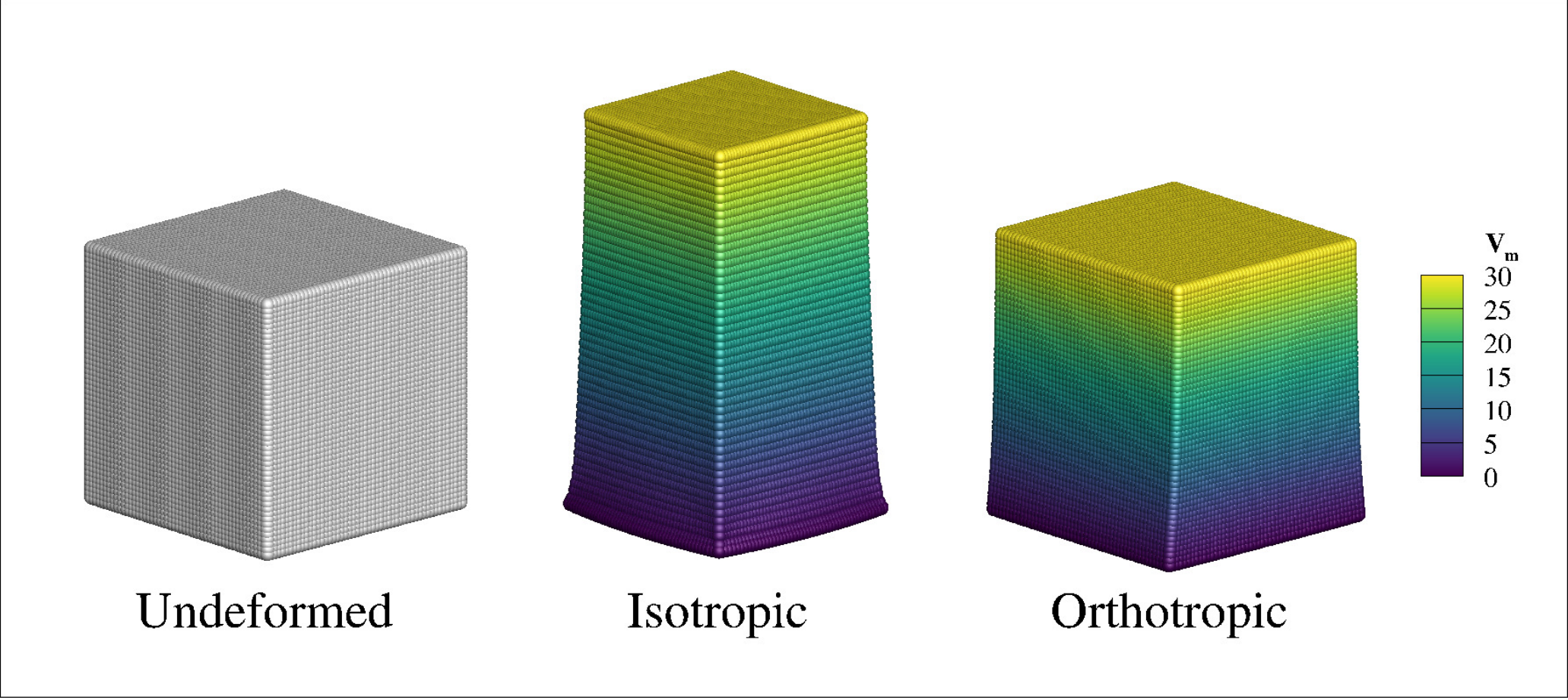}
		\caption{Active response of cubic myocardium.}
		\label{fig:active-muscle}
	\end{subfigure}
	\newline 
	\begin{subfigure}[b]{0.95\textwidth}
		\vspace{0.5cm}
		\includegraphics[trim = 1mm 1.5cm 1mm 2cm, clip, width=.975\textwidth]{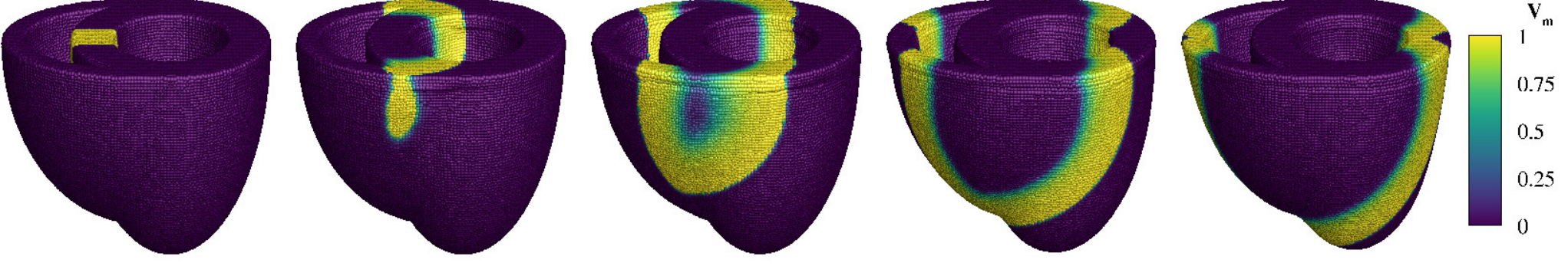}
		\includegraphics[trim = 1mm 2cm 1mm 2cm, clip, width=.975\textwidth]{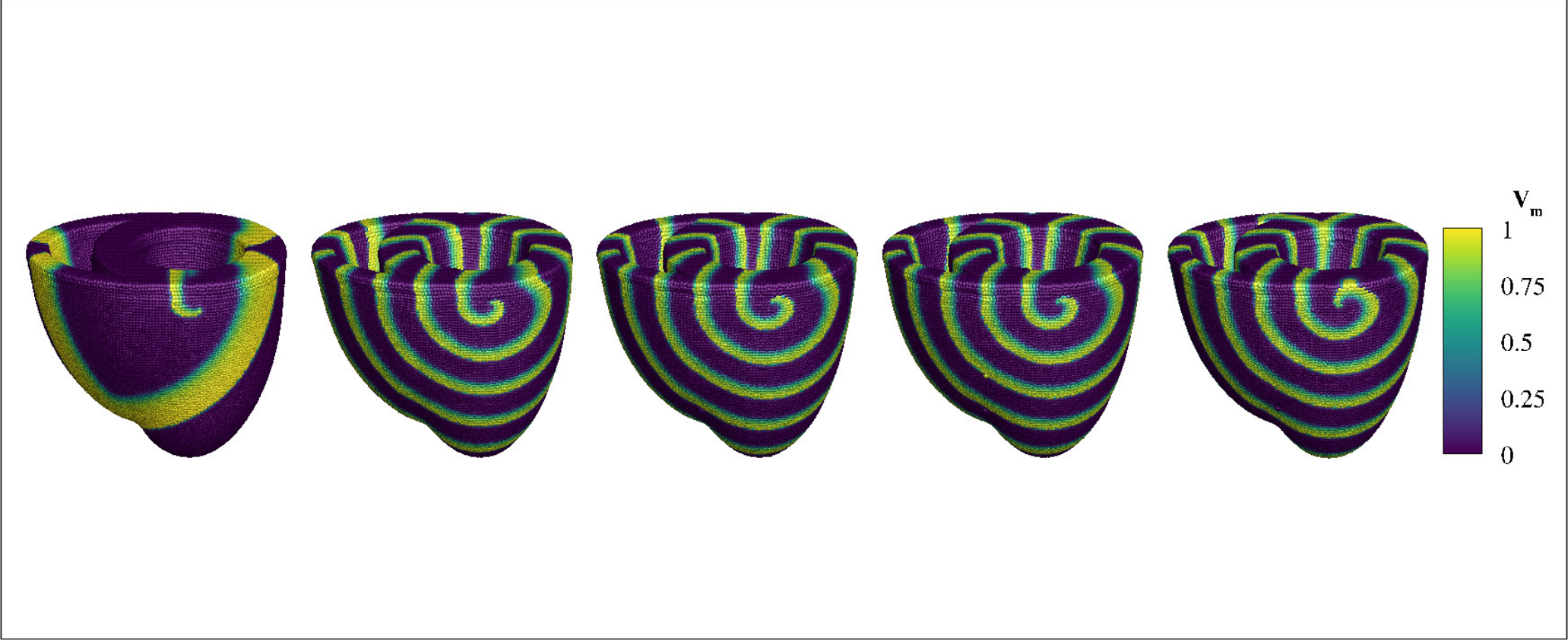}
		\caption{Excitation-contraction of generic biventricular heart.}
		\label{fig:biventricular-heart}
	\end{subfigure}
	\caption{SPHinXsys validations and applications for electro-mechanical coupling.
		(b) Active response of the unit cubic myocardium with both isotropic and anisotropic material properties.
		(c) Generic biventricular heart: coupled excitation-contraction induced by the transmembrane potential 
		as a free pulse (top panel) and scroll wave (bottom panel).}
	\label{fig:cardiacmodeling}
\end{figure*}

To validate the electromechanics solver, 
we consider a benchmark (folder "cases\_test/test\_3d\_active\_myocardium") where a unit cube of myocardium with 
a linearly distributed transmembrane potential, whose time variation is neglected. 
The electro-mechanical coupling is governed by a ad-hoc activation law and the constitutive law describing the passive response is the Holzapfel-Ogden model. 
Figure \ref{fig:active-muscle} shows the deformed configuration of the cubic myocardium and for the quantitative comparison with data in literature the reader is referred to Ref. \cite{zhang2020integrative}.
In this application (folder "cases\_test/test\_3d\_electro\_mechanics"), 
the SPHinXsys is applied for modeling excitation-induced contraction of a generic biventricular heart model. 
Two cases with free-pulse and scroll wave propagation of the transmembrane potential are considered. 
Figure \ref{fig:biventricular-heart} shows the resulting excitation-contraction of the heart with the transmembrane potential contours. 
%
%
%
\section{Concluding remarks and future work}\label{sec:conclusion}
SPHinXsys v0.2.0 is an open source SPH research library featured by several numerical schemes dealing with: 
fluid dynamics, solid mechanics, thermal and mass diffusion, 
reaction diffusion, fluid-structure interaction, electromechanics and their coupling with rigid body dynamics.
At present, 
the SPHinXsys major publications involve the validations and preliminary applications. 
SPHinXsys is developed and distributed on a GitHub public repository with comprehensive tutorials, 
thereby allowing the code availability and possible modification, 
along with the reproduciblity of the published test cases. 

In the future, 
one improvement would be the optimization the computational efficiency by implementing graphics processing unit (GPU) 
accelerators combined with many-core parallelization strategy. 

Another core aim is the development of an open heart simulator, 
which is expected to carry out numerical simulations of the total cardiac function \cite{quarteroni2017integrated}, 
based on the SPH method.  

In addition, more industrial applications, 
e.g. oscillating wave energy converter (OWSC),
will be added to the next version of SPHinXsys. 
%
%
\section*{CRediT authorship contribution statement}
{\bfseries  Chi Zhang:} Conceptualization, Methodology, Software (coding and testing of existing library components) , 
Validation, Formal analysis, Writing - original draft, Writing - review \& editing, Visualization. 
{\bfseries  Massoud Rezavand:} Methodology, Software (coding and testing of existing library components), Writing - review \& editing.
{\bfseries Yujie Zhu:} Methodology, Software (coding and testing of existing library components), Writing - review \& editing.  
{\bfseries Yongchuan Yu:} Methodology, Software (coding and testing of existing library components). 
{\bfseries Dong Wu:} Methodology, Software (coding and testing of existing library components). 
{\bfseries Wenbin Zhang:} Methodology, Software (coding and testing of existing library components). 
{\bfseries Jianhang Wang: } Methodology. 
{\bfseries Xiangyu Hu:} Supervision, Conceptualization, Methodology, Software (coding and overview), Writing - review \& editing .
%
%
\section*{Declaration of competing interest }
The authors declare that they have no known competing financial interests or personal relationships that could have appeared to influence the work reported in this paper.
%
%
\section{Acknowledgement}
The authors would like to express their gratitude to Deutsche Forschungsgemeinschaft 
for their sponsorship of this research under grant numbers 
DFG HU1527/10-1 and HU1527/12-1.
%
%
\section*{References}
\bibliography{mybibfile}

\begin{thebibliography}{10}
\expandafter\ifx\csname url\endcsname\relax
  \def\url#1{\texttt{#1}}\fi
\expandafter\ifx\csname urlprefix\endcsname\relax\def\urlprefix{URL }\fi
\expandafter\ifx\csname href\endcsname\relax
  \def\href#1#2{#2} \def\path#1{#1}\fi

\bibitem{zhang2017weakly}
C.~Zhang, X.~Hu, N.~A. Adams, A weakly compressible {SPH} method based on a
  low-dissipation riemann solver, J. Comput. Phys. 335 (2017) 605--620.

\bibitem{zhang2020dual}
C.~Zhang, M.~Rezavand, X.~Hu, Dual-criteria time stepping for weakly
  compressible smoothed particle hydrodynamics, Journal of Computational
  Physics 404 (2020) 109135.

\bibitem{zhang2019multi}
C.~Zhang, M.~Rezavand, X.~Hu, A multi-resolution sph method for fluid-structure
  interactions, arXiv preprint arXiv:1911.13255.

\bibitem{zhang2020integrative}
C.~Zhang, J.~Wang, M.~Rezavand, D.~Wu, X.~Hu, An integrative smoothed particle
  hydrodynamics framework for modeling cardiac function, arXiv preprint
  arXiv:2009.03759.

\bibitem{lucy1977numerical}
L.~B. Lucy, A numerical approach to the testing of the fission hypothesis, The
  Astronomical Journal 82 (1977) 1013--1024.

\bibitem{gingold1977smoothed}
R.~A. Gingold, J.~J. Monaghan, Smoothed particle hydrodynamics: theory and
  application to non-spherical stars, Mon. Not. R. Astron. Soc. 181~(3) (1977)
  375--389.

\bibitem{monaghan1994simulating}
J.~J. Monaghan, Simulating free surface flows with {SPH}, J. Comput. Phys.
  110~(2) (1994) 399--406.

\bibitem{hu2006multi}
X.~Hu, N.~Adams, A multi-phase {SPH} method for macroscopic and mesoscopic
  flows, J. Comput. Phys. 213 (2006) 844--861.

\bibitem{shao2006simulation}
S.~Shao, C.~Ji, D.~I. Graham, D.~E. Reeve, P.~W. James, A.~J. Chadwick,
  Simulation of wave overtopping by an incompressible {SPH} model, Coastal Eng.
  53~(9) (2006) 723--735.

\bibitem{zhang2019weakly}
C.~Zhang, G.~Xiang, B.~Wang, X.~Hu, N.~Adams, A weakly compressible {SPH}
  method with {WENO} reconstruction, Journal of Computational Physics 392
  (2019) 1--18.

\bibitem{libersky1991smooth}
L.~D. Libersky, A.~G. Petschek, Smooth particle hydrodynamics with strength of
  materials, in: Advances in the free-Lagrange method including contributions
  on adaptive gridding and the smooth particle hydrodynamics method, Springer,
  1991, pp. 248--257.

\bibitem{benz1995simulations}
W.~Benz, E.~Asphaug, Simulations of brittle solids using smooth particle
  hydrodynamics, Comput. Phys. Commun. 87~(1) (1995) 253--265.

\bibitem{monaghan2000sph}
J.~J. Monaghan, {SPH} without a tensile instability, J. Comput. Phys. 159~(2)
  (2000) 290--311.

\bibitem{randles1996smoothed}
P.~Randles, L.~Libersky, Smoothed particle hydrodynamics: some recent
  improvements and applications, Comput. Methods Appl. Mech. Eng. 139~(1-4)
  (1996) 375--408.

\bibitem{antoci2007numerical}
C.~Antoci, M.~Gallati, S.~Sibilla, Numerical simulation of fluid--structure
  interaction by {SPH}, Computers \& Structures 85~(11-14) (2007) 879--890.

\bibitem{han2018sph}
L.~Han, X.~Hu, {SPH} modeling of fluid-structure interaction, Journal of
  Hydrodynamics 30~(1) (2018) 62--69.

\bibitem{rezavand2020weakly}
M.~Rezavand, C.~Zhang, X.~Hu, A weakly compressible sph method for violent
  multi-phase flows with high density ratio, Journal of Computational Physics
  402 (2020) 109092.

\bibitem{morris1997modeling}
J.~P. Morris, P.~J. Fox, Y.~Zhu, Modeling low reynolds number incompressible
  flows using sph, J. Comput. Phys. 136~(1) (1997) 214--226.

\bibitem{Adami2013}
S.~Adami, X.~Y. Hu, N.~A. Adams, A transport-velocity formulation for smoothed
  particle hydrodynamics, J. Comput. Phys. 241 (2013) 292--307.

\bibitem{zhang2017generalized}
C.~Zhang, X.~Y. Hu, N.~A. Adams, A generalized transport-velocity formulation
  for smoothed particle hydrodynamics, J. Comput. Phys. 337 (2017) 216--232.

\bibitem{adami2012generalized}
S.~Adami, X.~Hu, N.~Adams, A generalized wall boundary condition for smoothed
  particle hydrodynamics, J. Comput. Phys. 231~(21) (2012) 7057--7075.

\bibitem{holzapfel2009constitutive}
G.~A. Holzapfel, R.~W. Ogden, Constitutive modelling of passive myocardium: a
  structurally based framework for material characterization, Philosophical
  Transactions of the Royal Society of London A: Mathematical, Physical and
  Engineering Sciences 367~(1902) (2009) 3445--3475.

\bibitem{vignjevic2006sph}
R.~Vignjevic, J.~R. Reveles, J.~Campbell, Sph in a total lagrangian formalism,
  CMC-Tech Science Press- 4~(3) (2006) 181.

\bibitem{tran2016simulation}
T.~Tran-Duc, E.~Bertevas, N.~Phan-Thien, B.~C. Khoo, Simulation of anisotropic
  diffusion processes in fluids with smoothed particle hydrodynamics,
  International Journal for Numerical Methods in Fluids 82~(11) (2016)
  730--747.

\bibitem{fitzhugh1961impulses}
R.~FitzHugh, Impulses and physiological states in theoretical models of nerve
  membrane, Biophys. J. 1~(6) (1961) 445.

\bibitem{quarteroni2017cardiovascular}
A.~Quarteroni, A.~Manzoni, C.~Vergara, The cardiovascular system: mathematical
  modelling, numerical algorithms and clinical applications, Acta Numerica 26
  (2017) 365--590.

\bibitem{aliev1996simple}
R.~R. Aliev, A.~V. Panfilov, A simple two-variable model of cardiac excitation,
  Chaos, Solitons \& Fractals 7~(3) (1996) 293--301.

\bibitem{panfilov1999three}
A.~Panfilov, Three-dimensional organization of electrical turbulence in the
  heart, Physical Review E 59~(6) (1999) R6251.

\bibitem{wang2019split}
J.-H. Wang, S.~Pan, X.~Y. Hu, N.~A. Adams, A split random time-stepping method
  for stiff and nonstiff detonation capturing, Combustion and Flame 204 (2019)
  397--413.

\bibitem{ten2004model}
K.~Ten~Tusscher, D.~Noble, P.-J. Noble, A.~V. Panfilov, A model for human
  ventricular tissue, American Journal of Physiology-Heart and Circulatory
  Physiology 286~(4) (2004) H1573--H1589.

\bibitem{nash2004electromechanical}
M.~P. Nash, A.~V. Panfilov, Electromechanical model of excitable tissue to
  study reentrant cardiac arrhythmias, Progress in biophysics and molecular
  biology 85~(2-3) (2004) 501--522.

\bibitem{wong2011computational}
J.~Wong, S.~G{\"o}ktepe, E.~Kuhl, Computational modeling of electrochemical
  coupling: a novel finite element approach towards ionic models for cardiac
  electrophysiology, Computer methods in applied mechanics and engineering
  200~(45-46) (2011) 3139--3158.

\bibitem{yang2014smoothed}
X.~Yang, M.~Liu, S.~Peng, Smoothed particle hydrodynamics modeling of viscous
  liquid drop without tensile instability, Computers \& Fluids 92 (2014)
  199--208.

\bibitem{wendland1995piecewise}
H.~Wendland, Piecewise polynomial, positive definite and compactly supported
  radial functions of minimal degree, Adv. Comput. Math. 4~(1) (1995) 389--396.

\bibitem{ratti2019}
L.~Ratti, M.~Verani, A posteriori error estimates for the monodomain model in
  cardiac electrophysiology, Calcolo 56.
\newblock \href {http://dx.doi.org/10.1007/s10092-019-0327-2}
  {\path{doi:10.1007/s10092-019-0327-2}}.

\bibitem{quarteroni2017integrated}
A.~Quarteroni, T.~Lassila, S.~Rossi, R.~Ruiz-Baier, Integrated heart—coupling
  multiscale and multiphysics models for the simulation of the cardiac
  function, Computer Methods in Applied Mechanics and Engineering 314 (2017)
  345--407.

\end{thebibliography}
%
%
\end{document}